\documentstyle[12pt,epsf]{article}
\def\hybrid{\topmargin 0pt      \oddsidemargin 0pt
        \headheight 0pt \headsep 0pt
        \textheight 9in         
        \textwidth 6.25in       
        \marginparwidth .875in
        \parskip 5pt plus 1pt   \jot = 1.5ex}

\catcode`\@=11
\def\marginnote#1{}
\newcount\hour
\newcount\minute
\newtoks\amorpm
\hour=\time\divide\hour by60
\minute=\time{\multiply\hour by60 \global\advance\minute by-\hour}
\edef\standardtime{{\ifnum\hour<12 \global\amorpm={am}%
        \else\global\amorpm={pm}\advance\hour by-12 \fi
        \ifnum\hour=0 \hour=12 \fi
        \number\hour:\ifnum\minute<10 0\fi\number\minute\the\amorpm}}
\edef\militarytime{\number\hour:\ifnum\minute<10 0\fi\number\minute}

\def\draftlabel#1{{\@bsphack\if@filesw {\let\thepage\relax
   \xdef\@gtempa{\write\@auxout{\string
      \newlabel{#1}{{\@currentlabel}{\thepage}}}}}\@gtempa
   \if@nobreak \ifvmode\nobreak\fi\fi\fi\@esphack}
        \gdef\@eqnlabel{#1}}
\def\@eqnlabel{}
\def\@vacuum{}
\def\draftmarginnote#1{\marginpar{\raggedright\scriptsize\tt#1}}

\def\draft{\oddsidemargin -.5truein
        \def\@oddfoot{\sl preliminary draft \hfil
        \rm\thepage\hfil\sl\today\quad\militarytime}
        \let\@evenfoot\@oddfoot \overfullrule 3pt
        \let\label=\draftlabel
        \let\marginnote=\draftmarginnote
   \def\@eqnnum{(\theequation)\rlap{\kern\marginparsep\tt\@eqnlabel}%
\global\let\@eqnlabel\@vacuum}  }

\def\numberbysection{\@addtoreset{equation}{section}
        \def\theequation{\thesection.\arabic{equation}}}

\def\underline#1{\relax\ifmmode\@@underline#1\else
        $\@@underline{\hbox{#1}}$\relax\fi}

\def\titlepage{\@restonecolfalse\if@twocolumn\@restonecoltrue\onecolumn
     \else \newpage \fi \thispagestyle{empty}\c@page\z@
        \def\thefootnote{\fnsymbol{footnote}} }

\def\endtitlepage{\if@restonecol\twocolumn \else  \fi
        \def\thefootnote{\arabic{footnote}}
        \setcounter{footnote}{0}}  
\catcode`@=12
\relax

\def\beq{\begin{equation}}
\def\eeq{\end{equation}}
\def\bea{\begin{eqnarray}}
\def\eea{\end{eqnarray}}
\def\nn{\nonumber}

\relax
\numberbysection
\hybrid
\begin{document}
\begin{titlepage}
\begin{center}
September~1997 \hfill    PAR--LPTHE 97/28 \\
               \hfill hep-th/9709136 \\[.2in]
{\large\bf Random Bond Potts Model: the Test of the Replica
Symmetry Breaking.}\\[.1in]

        {\bf Vik.S.Dotsenko$^1$, Vl.S.Dotsenko$^{1,2}$ and M. Picco$^{2,3}$}\\
        $^1${\it Landau Institute for Theoretical Physics,\\
        Russian Academy of Sciences,\\
        Kosygina 2, 117940  Moscow, Russia}\\
        $^2${\it LPTHE\/}\footnote{Laboratoire associ\'e No. 280 au CNRS}\\
       \it  Universit\'e Pierre et Marie Curie, PARIS VI\\
       \it Universit\'e Denis Diderot, PARIS VII\\
        Boite 126, Tour 16, 1$^{\it er}$ \'etage \\
        4 place Jussieu\\
        F-75252 Paris CEDEX 05, France\\
        $^3${\it Departamento de F\'{\i}sica,\\
        \it Universidade Federal do Esp\'{\i}rito Santo\\
        Vit\'oria - ES, Brazil\\}
        
\end{center}

\vskip .1in
\centerline{ABSTRACT}
\begin{quotation}
Averaged spin-spin correlation function squared
$\overline{<\sigma(0)\sigma(R)>^{2}}$ is calculated for the
ferromagnetic random bond Potts model. The technique being used is
the renormalization group plus conformal field theory. The results
are of the $\epsilon$ - expansion type fixed point calculation,
$\epsilon$ being the deviation of the central charge (or the number of
components) of the Potts model from the Ising model
value. Calculations are done both for the replica symmetric and the
replica symmetry broken fixed points. The results obtained allow for
the numerical simulation tests to decide between the two different
criticalities of the random bond Potts model.

\end{quotation}
\end{titlepage}
\newpage

\section{Introduction.} 

For the ferromagnetic random bond Potts model there exist two
different fixed point solutions, one which is replica symmetric [1,2]
and a second in which the replica symmetry is broken [3].  Both fixed
points are of a universal nature and one or another is reached
depending on the initial conditions for the coupling constant in the
renormalization group (RG) equation.
More specifically, if one starts with the model defined on a lattice,
then the partition function shall be given by:
\beq
Z(\beta)=\sum_{\{\sigma\}}\exp\{-\beta H[\sigma]\}
\eeq
where $H[\sigma]$ is a nearest neighbor interaction hamiltonian for
classical spins $\{\sigma_{x}\}$:
\beq
H[\sigma]=\sum_{x,\alpha}J_{x,\alpha}V(\sigma_{x},\sigma_{x+\alpha})
\eeq
Each spin $\sigma_{x}$ is taking $q$ values, in case of $q$-component
Potts model; $x$ stands for lattice sites, $(x,\alpha)$ - for lattice
bonds: in case of a square lattice $\alpha=1,2;\,\,
V(\sigma_{x},\sigma_{x+\alpha})$ is the spin-spin interaction
potential: in case of the Potts model the usual choice would be
$V(\sigma,\sigma') \propto
1-\delta_{\sigma,\sigma'}.\,\,\{J_{x,\alpha}\}$ are the bond coupling
constants. They are supposed to be ferromagnetic but taking random
values, independently for each lattice bond, with some distribution,
characterized by a width $g_{0}$, around a mean value $J_{0}$.

Weak disorder corresponds to a small value of $g_{0}:\,\,g_{0}\ll
J_{0}$. In this case the model could be studied in the continuum
limit, if $\beta$ is close to a critical value
$\beta_{c},\,\,(\beta-\beta_{c})/\beta_{c}=\tau\ll 1$. In this limit
the partition function could be given in the form:
\beq
Z(\beta)=Tr \exp\{-H_{0}-H_{1}\}
\eeq
$Tr$ stands symbolically to represent the sum over the spin
configurations, but in the context of the continuum limit theory. Its
explicit realization is not important because we shall eventually be
dealing with correlation functions, and these are defined
unambiguously by the corresponding conformal field theory.  $H_{0}$
stands for a hamiltonian, or a field theory action, of the conformal
field theory corresponding to a given $q$-states Potts model, being
defined on a perfect lattice with the spin-spin coupling constant
$J_{0}$ for all lattice bonds and taken at its critical point,
$\beta=\beta_{c};\,\,H_{1}$ represents a deviation from the critical
point and it contains disorder. $H_{1}$ could be given in the
following form:
\beq
H_{1}=\int d^{2}x\tau(x)\varepsilon(x)
\eeq
where
\beq
\tau(x)\propto\beta J(x)-\beta_{c}J_{0}
\eeq
is the random temperature parameter of the continuum limit theory; $x$
takes values on the continuum $2D$ plane; $J_{x,\alpha}$ of the
lattice is replaced by $J(x);\,\, \varepsilon(x)$ is the energy
operator of the Potts model replacing $V(\sigma_{x},
\sigma_{x+\alpha})$ on the lattice. As far as critical properties are
concerned the reduced continuum limit form of the model, defined by
the eqs.  (1.3), (1.4), is sufficient. For simplicity we shall assume
that $\tau(x)$ has a Gaussian distribution, for each $x$, with a width
$g_{0}$, so that
\beq
\overline{\tau(x)}=\tau_{0}=(\beta-\beta_{c})/\beta_{c}
\eeq
\beq
\overline{(\tau(x)-\tau_{0})(\tau(x')-\tau_{0})}=g_{0}\delta^{(2)}(x-x')
\eeq
The partition function (1.3) is of the form:
\beq
Z(\beta)=Tr \exp\{-H_{0}-\int d^{2}x\tau(x)\varepsilon(x)\}
\eeq
To take the average over the disorder one introduces replicas,
$n$ copies of the same model:
\beq 
(Z(\beta))^{n}=Tr\exp\{-\sum_{a=1}^{n}H_{0}^{(a)}-\int
d^{2}x\tau(x)\sum_{a=1}^{n} \varepsilon_{a}(x)\} 
\eeq
and then one takes the average:
\beq
\overline{(Z(\beta))^{n}}=Tr\exp\{-\sum_{a=1}^{n}H_{0}^{(a)}-
\tau_{0}\int d^{2}x\sum_{a=1}^{n}\varepsilon_{a}(x)+g_{0}\int 
d^{2}x\sum_{a\neq b}^{n}\varepsilon_{a}(x)\varepsilon_{b}(x)\}
\eeq
One arrives in this way at a homogeneous field theory of $n$ coupled
models with the coupling action:

\beq 
H_{\mbox{int}}=-g_{0}\int d^{2}x\sum_{a\neq
b}^{n}\varepsilon_{a}^{(x)}\varepsilon_{b}(x) 
\eeq
In $H_{\mbox{int}}$ the non-diagonal terms only, $a\neq b$, are being
kept. The diagonal ones could be put back into
$\sum_{a=1}^{n}H_{0}^{(a)}$. Moreover, in case of the Potts model the
energy operator $\varepsilon(x)$ corresponds to the operator
$\Phi_{1,2}(x)$ of the corresponding conformal theory. Its operator
algebra is known
\beq
\Phi_{1,2}\Phi_{1,2}\rightarrow\Phi_{1,1}+\Phi_{1,3}
\eeq
Here $\Phi_{1,1}=I$ is the identity operator, and the operator
$\Phi_{1,3}(x)$ is irrelevant, $2\Delta_{1,3}>2$. So there will be no
relevant diagonal subtractions from
$\sum_{a,b}\varepsilon_{a}\varepsilon_{b}$, the diagonal terms can
just be dropped.  In our analysis of the random bond Potts model we
shall take the replicated field theory form of it, equations (1.10),
(1.11), as our starting point.  In the renormalization group analysis,
with the interaction term given by the equation (1.11), the
qualitatively different initial conditions for the renormalization
group equation correspond to either taking the initial coupling
constant as it is in eq.(1.11), $g(\xi=0)=g_{0}$ ($\xi$ is the RG
parameter), or to breaking the replica symmetry initially by putting
$H_{\mbox{int}}$ into the form:
\beq 
H_{\mbox{int}}=-\int d^{2}x\sum_{a\neq
b}g_{ab}\varepsilon_{a}(x)\varepsilon_{b}(x) 
\eeq
This corresponds to assuming different couplings, initially, for
different replicas.

Replica symmetric form of $g_{ab}$ is
\beq
g_{ab}=g_{0},\,\,\,\mbox{all}\,\,\,a\neq b
\eeq
and $g_{ab}=0$ for $a=b$. Taking $g_{ab}$ with different components
corresponds to replica symmetry breaking perturbation. Relevance of
this type of perturbation has been observed initially in [4] for the
random $XY$ model and in [5] for the standard $\varphi^{4}$ theory,
with randomness. These two models move, under RG, into a strong
coupling regime. The first fixed point with a replica symmetry
breaking has been found in [3] for the 2D random-bond Potts model. For
the moment this remains to be the only solution of this type.

To repeat it again, the random bond Potts model, i.e. the model with
$H_{\mbox{int}}$ in (1.13), has two different fixed point solutions:

1) replica symmetric, if initially $g_{ab}$ of the form in eq.(1.14);

2) replica symmetry broken, if initially $g_{ab}$ is different from (1.14).

To check which one of the two solutions realizes for the original spin
model defined on a lattice with random bonds, eqs.(1.1), (1.2), in the
numerical simulation experiment in particular if not in the real one,
we shall calculate in this paper one particular observable quantity
for which the replica symmetry breaking effects are expected to be
most pronounced.  The calculations will be done for both RG fixed
points to make the verification problem well defined.

The quantity in question is the averaged spin-spin correlation function squared
\beq
\overline{<\sigma(0)\sigma(R)>^{2}}
\eeq
The related quantity is the renormalization group amplitude
\beq
Z_{ab}(\xi)
\eeq
for the replicated model operator
\beq
O_{ab}(x)=\sigma_{a}(x)\sigma_{b}(x)
\eeq
Here the spins $\sigma_{a},\,\,\sigma_{b}$, belong to different
replicas, $a\neq b$, and are taken at the same point. The amplitude
(1.16) is the spin-spin "overlap function" for our critical model, if
one uses the terminology of the theory of spin glasses [6].

Further discussions will be postponed until the formal analytic
solutions are obtained.

In the next section the analytic problem will be defined and the RG
equation for the amplitude $ Z_{ab}(\xi)$ will be obtained, up to
second order in the coupling constant $g_{ab}(\xi)$. In Section 3 the
amplitude $Z_{ab}$ and the correlation function
$\overline{<\sigma(0)\sigma(R)>^{2}}$ will be calculated. We shall
also define the corresponding magnetization.  In Section 4 the results
will be analyzed, in particular as applied to the 3 and 4 component
Potts models. We shall also consider in this section the distribution
function for products of local magnetizations. This distribution is
encoded in the spin-spin "overlap function", the amplitude $Z_{ab}$,
in a way analogous to that for the corresponding object in the
spin-glass theory.  In section 5 results of numerical
simulations will be presented for the 3 and 4 - component Potts
models.  Section 6 is devoted to conclusions and discussions.  In the
Appendices the calculations of particular complicated integrals are
exposed, which appear in the calculation of coefficients of the RG
equation for $Z_{ab}(\xi)$.

\section{Definition of the problem and calculation of the RG
equation for $Z_{ab}$.}

The RG equation for the coupling constant $g_{ab}$ in the action
$H_{\mbox{int}}$, eq.(1.13), has been derived and solved in [3], for
the replica symmetry broken fixed point. The replica symmetric
solution has originally been found in [1].  We shall use those results
but first we shall concentrate on the RG equation for the amplitude
$Z_{ab}(\xi)$. This function occurs in the calculation of the
correlation function $\overline{<\sigma(0)\sigma(R)>^{2}}.$

In terms of replicas the averaged spin-spin correlation function,
$\overline{\langle\sigma(0)\sigma(R)\rangle^{2}}\\ =\lim_{n\to
0}\frac{1}{n(n-1)}\sum_{a\neq b}
\langle\sigma_{a}(0)\sigma_{a}(R)\sigma_{b}(0)\sigma_{b}(R)\rangle$,
can be represented in the following form (see Sec.3, eqs.(3.7)-(3.9)):
\beq
\overline{<\sigma(0)\sigma(R)>^{2}}=\lim_{n\rightarrow 0}\frac{1}{2n(n-1)}
<\sum_{a\neq b}
\sigma_{a}(0)\sigma_{b}(0)\sum_{c\neq d}\sigma_{c}(R)\sigma_{d}(R)>
\eeq
Then the operator to be renormalized is:
\beq
O_{ab}(x)=\sigma_{a}(x)\sigma_{b}(x),\,\,\,a\neq b
\eeq
As we are going to use the perturbative RG, we have to consider 
the exponential of  $H_{\mbox{int}}$, 
the way it would 
enter the partition function (1.10), in the presence of the operator 
$O_{ab}(x)$:
\beq
O_{ab}(x)\exp\{-H_{\mbox{int}}\}=\sigma_{a}(x)\sigma_{b}(x)\exp\{\int d^{2}y
\sum_{a\neq b}g_{ab}\varepsilon_{a}(y)\varepsilon_{b}(y)\}
\eeq
We expand $\exp\{-H_{\mbox{int}}\}$:
\beq
O_{ab}(x)(1-H_{\mbox{int}}+\frac{1}{2}(H_{\mbox{int}})^{2}+\cdots)=O_{ab}(x)-O_{ab}(x)
H_{\mbox{int}}+\frac{1}{2}O_{ab}(x)H_{\mbox{int}}^{2}+\cdots
\eeq
and then we do all possible contractions and operator algebra, which
lead to reproducing the operator $O_{ab}(x)$. Contractions in the case
of our non-gaussian field theory amounts to equalizing replica indices
in all possible ways which occur under the summation over indices and
then replacing the products of operators with the same index by
correlation functions (or using directly the operator algebra) of the
unperturbed theory.  We remind that the unperturbed theory is a
collection of independent and identical Potts models. Each Potts
model, in the continuum limit, is a minimal conformal field theory
with the energy operator $\varepsilon(x)$ represented by the primary
field $\Phi_{1,2}(x)$, and with the central charge related to the
number of components $q$ of the original Potts model as defined on the
lattice. We shall specify this relation later, see also [7].

Proceeding in this way, one gets in the first order:
\bea
-O_{ab}(x)H_{\mbox{int}}&=&\sigma_{a}(x)\sigma_{b}(x)
\int d^{2}y\sum_{c\neq d} g_{cd}\varepsilon_{c}
(y)\varepsilon_{d}(y)\nn\\&&\rightarrow
2g_{ab}\sigma_{a}(x)\sigma_{b}(x)\int_{1<|y-x|<a}
d^{2}y\frac{D^{2}}{|x-y|^{2\Delta_{\epsilon}}}
\eea
We have contracted here the indices 
$b$ and $c$, and $a$ and $d$ (or $b$ and $d$, and $a$ 
and $c$, which amounts to a combinatorial 
factor of 2), and then we have used the operator algebra:
\beq
\sigma(x)\varepsilon(y)=\frac{D}{|x-y|^{\Delta_{\varepsilon}}}\sigma(x)
+\cdots
\eeq
$D$ stands for the operator algebra coefficient
$D^{\sigma}_{\sigma,\varepsilon}$.  For minimal conformal theories
these coefficients have been calculated in [8].

Integration in (2.5) is around $x$, over the distances ranging between
the "old" and the "new" cut-offs. The old one we choose to be equal to
1 (like it would be on a lattice, when distances are measured in
lattice spacings) and the new one $a,\,\,a>>1$.

$\varepsilon(x)$ is the primary field $\Phi_{1,2}$, so one has:
\beq
\Delta_{\varepsilon}=\Delta_{1,2}+\bar{\Delta}_{1,2}=2\Delta_{1,2}
\eeq
We use next the Kac formula for conformal dimensions of primary fields
$\Phi_{n',n}(x)$:
\beq
\Delta_{n',n}=\frac{(\alpha_{n'}+\alpha_{+}n)^{2}-(\alpha_{-}+\alpha_{-})^{2}}{4}
\eeq
The parameters $\alpha_{+},\alpha_{-}$ are related to the central charge of the 
Virasoro algebra:
\bea
c&=&1-24\alpha_{0}^{2},\nn\\
\alpha_{\pm}&=&\alpha_{0}\pm\sqrt{1+\alpha_{1}^{2}}
\eea
Notice that
\beq
\alpha_{+}\alpha_{-}=-1
\eeq
We shall use in the following the parameter $\alpha_{+}^{2}$ to
specify the models, instead of the central charge $c$. For Ising model
$c=\frac{1}{2},\,\,\alpha^{2}_{+}=\frac{4}{3}$. One gets conformal
theories for the critical Potts models with the number of components
$q$ varying continuously from 2 to 4 if the central charge is taken to
vary between 1/2 and 1, or $\alpha_{+}^{2}$ varying between 4/3 and 1
[7]. For the Ising model
$\Delta_{1,2}=\frac{1}{2},\,\,\Delta_{\varepsilon}=1$, and the
perturbation $H_{\mbox{int}}$ (1.13) is marginal, $g_{ab}$ is
dimensionless. This perturbation becomes slightly relevant if,
following Ludwig [1], one takes
\beq
\alpha_{+}^{2}=\frac{4}{3}-\epsilon
\eeq
and one studies the Potts models by the $\epsilon$-expansion RG
assuming $\epsilon$ to be small.

The dimension of the energy operator is now given by:
\beq
\Delta_{\varepsilon}=2\Delta_{1,2}=\frac{(\alpha_{-}+2\alpha_{+})^{2}-(\alpha_{-}+
\alpha_{+})^{2}}{2}=1-\frac{3}{2}\epsilon
\eeq
The first order correction in (2.5) takes the form:
\beq
2g_{ab}\sigma_{a}(x)\sigma_{b}(x)D^{2}
\int_{1<|y|<a}\frac{d^{2}y}{|y|^{2-3\epsilon}}=
2g_{ab}\sigma_{a}(x)\sigma_{b}(x)D^{2}2\pi\frac{1}{3\epsilon}a^{3\epsilon} 
\eeq
In integrating over scales from 1 to $a$ one gets a factor
$\frac{1}{\epsilon}(a^{3\epsilon}-1)$. We have replaced it, in a
standard way, by $\frac{1}{\epsilon}a^{3\epsilon}$, assuming $
a^{3\epsilon}\gg 1$.

Next, if one introduces the amplitude $Z_{ab}$ and studies
renormalization of the operator
\beq
\tilde O_{ab}(x) = Z_{ab}\sigma_{a}(x)\sigma_{b}(x)
\eeq
then (2.13) will correspond to the first order correction to $Z_{ab}$:
\beq
\delta Z_{ab}^{(1)}=Z_{ab}4\pi D^{2}g_{ab}\frac{1}{3\epsilon}a^{3\epsilon}
\eeq
This correction, the corresponding RG equation, and the renormalized
amplitude $Z_{ab}$ has first been defined in [1], (for the replica
symmetric case), as well as the corresponding amplitudes for higher
moments of $\sigma$: $\tilde O_{ab\cdots d} =
Z_{ab\cdots d}\sigma_{a}\sigma_{b}\cdots\sigma_{d}, a\neq b\neq\cdots\neq d.$

It has been shown in [3] that the replica symmetry breaking effects
appear in the second order. One needs at least two orders, for all the
quantities, to treat these effects. For this reason our analytic
problem will be to define renormalization of $Z_{ab}$ in two orders.

From eqs.(2.3), (2.4) one gets in the second order:
\bea
O_{ab}\frac{1}{2}(H_{\mbox{int}})^{2}&=&Z_{ab}\sigma_{a}(x)\sigma_{b}(x)\frac{1}{2}
\int d^{2}y\sum_{c\neq d} g_{cd}\varepsilon_{c}(y)\varepsilon_{d}(y)
\int d^{2}y'\sum_{e\neq f}g_{ef}\varepsilon_{e}(y')\varepsilon_{f}(y')\nn\\&&
\rightarrow
D_{1}^{(2)}+D_{2}^{(2)}+D_{3}^{(2)}
\eea
We have defined as $D^{(2)}_{i}$ the diagrams of the second order. 
The first diagram is of the form:
\bea
D^{(2)}_{1}&=& 8Z_{ab}\sigma_{a}(0)\sigma_{b}(0)\frac{1}{2}\sum_{d}g_{bd}g_{ad}\nn\\
&&\times \int d^{2}y\int d^{2}y'<\sigma(0)\varepsilon(y)\sigma(\infty)> 
<\sigma(0)\varepsilon(y')\sigma(\infty)>
<\varepsilon(y)\varepsilon(y')>\nn\\
&=&4Z_{ab}\sigma_{a}(0)\sigma_{b}(0)(g^{2})_{ab}
\int d^{2}y\int d^{2}y'\frac{D}{|y|^{\Delta_{\varepsilon}}}
\frac{D}{|y'|^{\Delta_{\varepsilon}}}\frac{1}{|y-y'|^{2\Delta_{\varepsilon}}}
\eea
To simplify somewhat the expressions in the integrals we have put the
operator $\tilde O_{ab}(x) = Z_{ab}\sigma_{a}(x)\sigma_{b}(x)$ at
$x=0$.

To get $D^{(2)}_{1}$ we have made equal in (2.16) $b=c$ and $a=e$ (we
remind that $a\neq b$, $c\neq d,\,\,e\neq f:$ diagonal elements of
$Z_{ab}, g_{cd},\,\,g_{ef}$ are assumed to be zero). Next we have used
the operator product expansion (2.6) for the products
$\sigma(0)\varepsilon(y)$ and $\sigma(0)\varepsilon(y')$, keeping just
the first term, the only relevant one, or, which is the same, we have
calculated the projections of $\sigma(0)\varepsilon(y)$ and of
$\sigma(0) \varepsilon(y')$ on the spin operator placed at
infinity. In addition, we have assumed the replica coupling matrix
$g_{cd}$ to be symmetric, $g_{cd}=g_{dc}$, as this is the case for the
Parisi matrices and for the fixed point solution for $g_{ab}$ in
[3]. In this case $\sum_{d}g_{bd}g_{da}=
\sum_{d}g_{ad}g_{db}=(g^{2})_{ab}$. Finally, the extra factor of 8 in
the first line in (2.17) is due to combinatorics: there are 8 ways to
make coupling of indices which give the diagram $D_1^{(2)}.$

The calculation of the integral in (2.17) is straightforward, it is
exposed in the Appendix A.1. One gets the following result: 
\beq
D^{(2)}_{1}=Z_{ab}\sigma_{a}(0)\sigma_{b}(0)8\pi^{2}D^{2}(g^{2})_{ab}(1+\epsilon
K)\frac{1} {9\epsilon^{2}}a^{6\epsilon}+O(1) 
\eeq 
where $K=6\log2$. In (2.18) we have kept only the terms which are 
singular in $\epsilon$, $\sim\frac{1}{\epsilon^{2}}$ and
$\frac{1}{\epsilon}$, the ones which are relevant for the RG.

The next diagram is obtained from (2.16) by setting $b=c=e$ and,
separately, $d=f$, plus all equivalents. This gives: 
\beq
D^{(2)}_{2}\propto Z_{ab}\sigma_{a}(0)\sigma_{b}(0)(g^{2})_{bb} \int
d^{2}y\int d^{2}y'
<\sigma(0)\varepsilon(y)\varepsilon(y')\sigma(\infty)>
<\varepsilon(y)\varepsilon(y')> 
\eeq
This diagram does not produce singularities in $\epsilon $, it is
finite for $\epsilon\rightarrow 0$, see Appendix A.2. As a result, it
does not contribute to the RG evolution of $Z_{ab}$.

Next diagram, $D^{(2)}_{3}$, is obtained by setting $b=c=e$ and
$a=d=f$, plus all equivalents. This gives the following expression:
\beq
D^{(2)}_{3}=4 Z_{ab}\sigma_{a}(0)\sigma_{b}(0)\frac{1}{2}(g_{ab})^{2}\int
d^{2}y\int
d^{2}y'(<\sigma(0)\varepsilon(y)\varepsilon(y')\sigma(\infty)>)^{2}
\eeq 
Calculation of this integral, which is fairly complicated, is
described in the Appendices A and B. Finally one gets the following
result: 
\beq 
D^{(2)}_{3}=
Z_{ab}\sigma_{a}(0)\sigma_{b}(0)(g_{ab})^{2}(8\pi^{2}D^{4}\frac{1}{9\epsilon^{2}}
-\frac{\pi^{2}}{3}\frac{1}{6\epsilon})a^{6\epsilon} 
\eeq 
Putting
(2.18) and (2.21), for $D^{(2)}_{1}$ and $D^{(2)}_{3}$, into (2.16),
and dropping $D^{(2)}_{2}$, eq.(2.19), one derives the second order
correction to the amplitude $Z_{ab}$: 
\beq
\delta
Z^{(2)}_{ab}=Z_{ab}\{8\pi^{2}D^{2}(1+\epsilon
K)(g^{2})_{ab}\frac{1}{9\epsilon^{2}}
+8\pi^{2}D^{4}(g_{ab})^{2}\frac{a^{6\epsilon}}{9\epsilon^{2}}
-\frac{\pi^{2}}{3}(g_{ab})^{2}
\frac{1}{6\epsilon}\}a^{6\epsilon} 
\eeq
Together with $\delta
Z^{(1)}_{ab}$ in (2.15) one obtains, up to second order: 
\bea
\tilde{Z}_{ab}&=&Z_{ab}+\delta Z^{(1)}_{ab}+\delta
Z^{(2)}_{ab}\nn\\&&=Z_{ab} \{1+4\pi D^{2}g_{ab}\frac{1}{
3\epsilon}a^{3\epsilon} + 8\pi D^{2}(1+\epsilon
K)(g^{2})_{ab}\frac{1}{9\epsilon^{2}}a^{6\epsilon} \nn \\
&& \qquad \quad +8\pi^{2}D^{4}(g_{ab})^{2}\frac{1}{9\epsilon^{2}}a^{6\epsilon}
-\frac{\pi^{2}} {3}(g_{ab})^{2}\frac{1}{6\epsilon}a^{6\epsilon}\} 
\eea 
Further in this Section we shall denote by tilde, like
$\tilde{Z}_{ab}$ or $\tilde g_{ab}$, the corresponding renormalized
quantities.

One obtains, in a standard way, the RG equation for $\tilde{Z}_{ab}$
by taking a derivative with respect to $\xi=\log a$:
\bea
\frac{d\tilde{Z}_{ab}}{d\xi}&\equiv & a\frac{d\tilde{Z}_{ab}}{da}\nn\\
&=& Z_{ab}\{4\pi D^{2} g_{ab}a^{3\epsilon}
+8\pi^{2}D^{2}(1+\epsilon K)(g^{2})_{ab}\frac{2}{3\epsilon}a^{6\epsilon}
\nn\\
&&\qquad \quad +8\pi^{2}D^{4}(g_{ab})^{2}\frac{2}{3\epsilon}a^{6\epsilon}
-\frac{\pi^{2}}{3}(g_{ab})^{2} a^{6\epsilon}\}
\eea
In the r.h.s. of this equation we have to replace $Z_{ab}$ and
$g_{ab}$ by the corresponding quantities renormalized in the first
order. One easily checks, by using the technique described above, and,
in addition, in case of $g_{ab}$, a dilatation of coordinates, in
order to return to the cut-off scale $a=1$, that to the first order:
\beq
\tilde{g}_{ab}=a^{3\epsilon}(g_{ab}+4\pi(g^{2})_{ab}\frac{a^{3\epsilon}}{3\epsilon})
\eeq
\beq
\tilde{Z}_{ab}=Z_{ab}(1+4\pi D^{2}g_{ab}\frac{a^{3\epsilon}}{3\epsilon})
\eeq

We have not been adding the dilatation term in case of renormalization
of $Z_{ab}$ renormalization because, by its usual definition, we
have put in this amplitude only the terms which are produced by
interactions. The trivial scaling factor in the renormalization of the
operator $O_{ab}(x)$ will be supplied later when we shall be
calculating the correlation function.

Inversely, in the first order:
\beq
g_{ab}=a^{-3\epsilon}(\tilde{g}_{ab}-4\pi D^{2}(\tilde{g}^{2})_{ab}
\frac{1}{3\epsilon})
\eeq
\beq
Z_{ab}=\tilde{Z}_{ab}(1-4\pi D^{2}\tilde{g}_{ab}\frac{1}{3\epsilon})
\eeq
Substituting these expressions in the r.h.s. of (2.24), 
and keeping terms up to second order
in $g_{ab}$, one obtains:
\beq
\frac{d\tilde{Z}_{ab}}{d\xi}=\tilde{Z}_{ab}\{4\pi D^{2}\tilde{g}_{ab}
+\frac{16\pi^{2}}{3}
D^{2}K(\tilde{g}^{2})_{ab}-\frac{\pi^{2}}{3}(\tilde{g}_{ab})^{2}\}
\eeq
This is the RG equation for the amplitude $\tilde{Z}_{ab}$ that we were aiming at.

It is easy to check that in the replica symmetric case:
\beq
g_{ab}\rightarrow g
\eeq
\beq
Z_{ab}\sigma_{a}\sigma_{b}\rightarrow Z\sigma_{a}\sigma_{b}
\eeq
-- for all $a,b$ $(a\neq b)$, the RG equation (2.29) takes the form:
\beq
\frac{d\tilde{Z}}{d\xi}=\tilde{Z}\{4\pi D^{2}\tilde{g}
+\frac{16\pi^{2}}{3}D^{2}K\tilde{g}^{2}
(n-2)-\frac{\pi^{2}}{3}\tilde{g}^{2}\}
\eeq
Here $n$ is the number of replicas, to be put equal to 0 eventually.

In the next Section we shall derive solutions for both equations,
(2.29) and (2.32), in order to have predictions for both fixed points,
symmetric and non-symmetric one.

\section{Solution of the RG equation for the amplitude $Z_{ab}$. 
Correlation function $\overline{<\sigma(0)\sigma(R)>^{2}.}$}

We shall drop tildes in $\tilde{g}_{ab}$ and $\tilde{Z}_{ab}$ 
in the following, as all the quantities that we shall use will be 
the renormalized ones. Also, to simplify
somewhat the equations, we shall change the normalization of $g_{ab}$:
\beq
g_{ab}\rightarrow\frac{g_{ab}}{4\pi}
\eeq
And we shall substitute the value of the operator algebra constant:
\beq
D=D(\epsilon)\equiv D^{\sigma}_{\sigma,\varepsilon}(\epsilon)=\frac{1}{2}+O(\epsilon^{2})
\eeq
It is well known that for the Ising model  $D^{\sigma}_{\sigma,\epsilon}=1/2$. One could check
that the correction is $\sim\epsilon^{2}$, by using the 
formulas derived in [8]. For our present 
calculation the $\epsilon^{2}$ correction is
irrelevant, so we can put the Ising model value $D=1/2.$

The equation for $Z_{ab}$ (2.29) takes the following form:
\beq
\frac{dZ_{ab}(\xi)}{d\xi}=Z_{ab}(\xi)\gamma_{ab}(\xi)
\eeq
\beq
\gamma_{ab}(\xi)=\frac{1}{4}g_{ab}(\xi)+\frac{K}{12}(g^{2}(\xi))_{ab}
-\frac{1}{48}(g_{ab}(\xi))^{2}
\eeq
We remind that the constant $K=6\log 2$. For the replica symmetric case we 
shall have:
\beq
\frac{dZ(\xi)}{d\xi}=Z(\xi)\gamma(\xi)
\eeq
\beq
\gamma(\xi)=\frac{1}{4}g(\xi)+\frac{K}{12}g^{2}(\xi)(n-2)
-\frac{1}{48}g^{2}(\xi)
\eeq

We are interested in the correlation function 
$\overline{<\sigma(0)\sigma(R)>^{2}}$ which is expressed through 
replicas by eq.(2.1). Assuming the RG
evolution from the lattice cut-off up to the scale $\sim R$, one gets:
\bea
\overline{<\sigma(0)\sigma(R)>^{2}} &\sim&\lim_{n\rightarrow 0}\frac{1}{2n
(n-1)}\sum_{a\neq b}\sum_{c\neq d}Z_{ab}(\xi_{R})Z_{cd}(\xi_{R})\frac{1}{R^{4
\Delta^{(0)_{\sigma}}}}\nn\\ 
&&\times <\sigma_{a}(0)\sigma_{b}(0)\sigma_{c}(1)\sigma_{d}(1)>
\eea
Here $\xi_{R}=\log R;\,\,\Delta^{(0)}_{\sigma}$ is the unperturbed dimension
of the spin operator in the Potts model. 
The factor $1/R^{2\Delta^{(0)}_{\sigma}}$ could have been
a part of $Z_{ab}(\xi)$. But, unlike for $g_{ab}$, we have included in $Z_{ab}$
only the scaling effects due to interactions. In this case the factor due to
trivial scaling of the unperturbed operator 
$\sigma$ has to be added separately.

Correlation of spins at the cut-off distance $a\sim 1$ is
that of the unperturbed decoupled replica models: 
it is $\sim 1$ if pairs of spins
have same replica indices and $0$ otherwise:
\beq
<\sigma_{a}(0)\sigma_{b}(0)\sigma_{c}(1)\sigma_{d}(1)>\sim\delta_{ac}\delta_{bd}
+\delta_{ad}\delta_{bc}
\eeq
One gets from (3.7):
\beq
\overline{<\sigma(0)\sigma(R)>^{2}}\sim\lim_{n\rightarrow 0}\frac{1}{n(n-1)}\sum_{
a\neq b}(Z_{ab}(\xi_{R}))^{2}\frac{1}{R^{4\Delta_{\sigma}^{(0)}}}
\eeq
The matrix $Z_{ab}(\xi)$ is assumed to be symmetric in its indices.

To find $\overline{<\sigma(0)\sigma(R)>^{2}}$, it remains to define $Z_{ab}(\xi)$. We
shall do it first for the replica symmetric fixed point, which is simpler.

\underline{RS.} In this case $Z_{ab}=Z,\,\,g_{ab}=g,\,\,a\neq b.$ Equation (3.9)
takes the form:
\beq
\overline{<\sigma(0)\sigma(R)>^{2}}\sim(Z(\xi_{R}))^{2}\frac{1}{R^{4\Delta^{(0)}
_{\sigma}}}
\eeq
$Z(\xi)$ could be defined from the equations (3.5), (3.6). We shall be 
deriving correlations for the model which is assumed to be already at 
the fixed point, and not for the crossover behavior
when the fixed point is approached. (In any case, for the replica 
symmetry broken case we have solution for $g_{ab}$ for the fixed point 
only.) In this case we have to substitute in eq.(3.6) the fixed point 
value of $g$. This has been derived in [1], up to second order
in $\epsilon $, and reproduced, by a somewhat different technique, 
in [2]. In the normalization that we have chosen at the start of 
this section, eq.(3.1), the fixed point value of $g$ is given by:
\beq
g_{*}=\frac{3}{2}\epsilon+\frac{9}{4}\epsilon^{2}+O(\epsilon^{3})
\eeq
From eqs.(3.5), (3.6), for $n=0$, we obtain:
\beq
\gamma_{*}=\frac{1}{4}g_{*}-(\frac{K}{6}+\frac{1}{48})g_{*}^{2}=\frac{3}{8}\epsilon-(\frac
{9}{4}\log 2-\frac{33}{64})\epsilon^{2}+0(\epsilon^{3})
\eeq
\beq
Z(\xi_{R})\sim\exp\{\gamma_{*}\xi_{R}\}=(R)^{\gamma_{*}}
\eeq
We have substituted $K=6\log 2$.
From (3.10):
\beq
\overline{<\sigma(0)\sigma(R)>^{2}}\sim\frac{1}{(R)^{2\Delta'_{\sigma^{2}}}}
\eeq
with 
\beq
\Delta'_{\sigma^{2}}=2\Delta_{\sigma}^{(0)}-\gamma_{*}
\eeq
By $\Delta'_{\sigma^{2}}$ we have denoted the scaling dimension of the operator
\beq
O_{ab}(x)=\sigma_{a}(x)\sigma_{b}(x),\,\,a\neq b
\eeq
at the replica symmetric fixed point.

\underline{RSB.} In the replica symmetry broken case all the matrices are assumed
to be in Parisi block-diagonal form. One passes to the continuous dependence
on matrix indices by using the following rules [6]:
\beq
g_{ab}\rightarrow g(t)
\eeq
\beq
(g^{2})_{aa}\rightarrow -\int^{1}_{0}dt g^{2}(t)
\eeq
\beq
 (g^{2})_{ab}\rightarrow -2g(t)\bar{g}-\int_{0}^{t}du(g(t)-g(u))^{2}
\eeq
with
\beq
\bar{g}=\int_{0}^{1}dtg(t)
\eeq
and similar expressions for the matrices $Z_{ab}$ and
$\gamma_{ab}$. The continuous parameter $t$ replaces indices, the way
described in [6]. It varies originally in the range from $1$ to $n$,
natural for the matrix indices, but after the limit $n\rightarrow 0$
is taken the interval of values of $t$ becomes [0,1]. The expressions
(3.17)- (3.19) are somewhat special since we assume that diagonal
elements of $g_{ab}$ (and of $Z_{ab}, \,\,\gamma_{ab}$) are zero. Note
also that there is no summation aver the index $a$ in (3.18), and that
for the Parisi matrices all the diagonal elements of the matrix
$g^{2}$ are equal, $a$- independent.

Using these rules one gets for $\overline{<\sigma(0)\sigma(R)>^{2}}$ from (3.9):
\bea
<\sigma(0)\sigma(R)>^{2}&\sim &\lim_{n\rightarrow 0}\frac{1}{n(n-1)}\sum_{a=1}^{n}
(Z^{2}(\xi_{R}))_{aa}\frac{1}{R^{4\Delta_{\sigma}^{(0)}}}\nn\\
&=&\lim_{n\rightarrow 0}\frac{1}{n-1}(Z^{2}(\xi_{R}))_{aa}
\frac{1}{(R)^{4\Delta_{\sigma}^{(0)}}}\nn\\
&=&\int_{0}^{1}dt Z^{2}(\xi_{R},t)\frac{1}{R^{4\Delta_{\sigma}^{(0)}}}
\eea
The equations (3.3) and (3.4) for $Z_{ab}$ and $\gamma_{ab}$ take on the following forms:
\beq
\frac{dZ(\xi,t)}{d\xi}=Z(\xi,t)\gamma(t)
\eeq
\beq
\gamma(t)=\frac{1}{4}g(t)+
\frac{K}{12}(-2g(t)\bar{g}-\int_{0}^{t}du(g(t)-g(u))^{2})-
\frac{1}{48}g^{2}(t)
\eeq
We have assumed here that the model is at the fixed point, so that $g_{ab}$ and
$\gamma_{ab}$ are $\xi $ independent. $g(t)$ is a fixed point function of $t$ only.
It has been found in [3] and it has the following form:
\beq
g(t)=\cases{\frac{1}{3}t,\,\,\,0<t<t_{1}\cr g_{1},\,\,\,t_{1}<t<1\cr}
\eeq
\beq
g_{1}=\frac{3}{2}\epsilon+\frac{9}{2}\epsilon^{2}+O(\epsilon^{3})
\eeq
\beq
t_{1}=3g_{1}
\eeq
We notice that $g(t)$ has a linearly growing piece for $0<t<t_{1}$, 
which makes it different from the replica symmetric solution:
\beq
g_{r.s.}(t) = \mbox{const} = g_{*}
\eeq
$g_{*}$ is given by eq.(3.11). We observe that $g_{*}$ is different in
its $\epsilon^{2}$ term from the constant part of $g(t),\,\,g_{1}$, eq.(3.25).

We notice also that the form of $g(t)$ in (3.24) 
corresponds to the full replica
symmetry breaking, in the terminology of the spin-glass theory [6]. 

$\gamma(t)$ is
defined by the eq.(3.23). It could still be simplified. First, from (3.20), (3.23):
\beq
\bar{g}=\int_{0}^{1}dtg(t)=g_{1}-\frac{3}{2}g^{2}_{1}
\eeq
Keeping the second term of $\bar{g}$ in the product $g(t)\bar{g}$ 
in (3.23) would
mean having terms $\sim g_{1}^{3}\sim\epsilon^{3}$ in this equation, which
exceeds the accuracy of our calculations. Second, the integral term $\int_{0}^{t}
du(g(t)-g(u))^{2}$ is $\sim g_{1}^{3}\sim\epsilon^{3}$. So it can also be dropped.
Then $\gamma(t)$ takes the following form:
\beq
\gamma(t)\simeq(\frac{1}{4}-\frac{K}{6}g_{1})g(t)-\frac{1}{48}g^{2}(t)=
a_{1}g(t)-a_{2}g^{2}(t)
\eeq
where
\beq
a_{1}=\frac{1}{4}-g_{1}\log 2 
\simeq\frac{1}{4}-\frac{3}{2}\epsilon\log 2
\eeq
\beq
a_{2}=\frac{1}{48}
\eeq
From eq.(3.22)
\beq
Z(\xi,t)\sim\exp\{\xi\gamma(t)\}
\eeq
For the correlation function one obtains from (3.21):
\bea
\overline{<\sigma(0)\sigma(R)>^{2}}&\sim&\int^{1}_{0}dt 
Z^{2}(\xi_{R},t)\frac{1}{R^{4\Delta^{(0)}_{\sigma}}}\nn\\
&=&\int_{0}^{1}dt\exp\{2\xi_{R}\gamma(t)\}\frac{1}{R^{4\Delta^{(0)}_{\sigma}}}
\eea
Here $\xi_{R}=\log R$. $2\xi_{R}\gamma(t)$ is typically small, $<1$.
This is the case for example 
for the 3-component Potts model and $R\sim 10^{3}$, the
biggest lattice sizes accessible for numerical simulation experiments. 
In this case it is
reasonable to calculate the integral in (3.33) by expanding the exponent. 
In this way one obtains:
\bea
R^{4\Delta_{\sigma}^{(0)}}\overline{<\sigma(0)\sigma(R)>^{2}}&\sim &
\int_{0}^{1}dt (1+2\gamma(t)\xi_{R}+2\gamma^{2}(t)\xi^{2}_{R}+\cdots)\nn\\
&=&1+2\bar{\gamma}\xi_{R}+ 2\overline{\gamma^{2}}\xi^{2}_{R}+\cdots
\eea
Here
\beq
\bar{\gamma}=\int_{0}^{1}dt\gamma(t)\simeq a_{1}\bar{g}-a_{2}\overline{g^{2}}
\eeq
\beq
\overline{\gamma^{2}}=\int^{1}_{0}dt\gamma^{2}(t)\simeq a_{1}^{2}\overline{g^{2}}-
2a_{1}a_{2}\overline{g^{3}}
\eeq
We remind that the precision of our calculations allows us to keep only the 
two first terms in $\bar{\gamma}$ or in $\overline{\gamma^{2}}$.
For
$\bar{g}, \overline{g^{2}}, \overline{g^{3}}$ one finds:
\beq
\bar{g}\simeq g_{1}-\frac{3}{2}g_{1}^{2}
\eeq
\beq
\overline{g^{2}}\simeq g_{1}^{2}-2g^{3}_{1}
\eeq
\beq
\overline{g^{3}}\simeq g^{3}_{1}-\frac{9}{4}g_{1}^{4}
\eeq
From (3.35), (3.36), (3.30)--(3.31), (3.37)--(3.39) one obtains:
\bea
\bar{\gamma}&\simeq & \frac{1}{4}g_{1}-(\log 2+\frac{19}{48})g^{2}_{1}\nn\\
&=& \frac{3}{8} \epsilon -(\frac{9}{4}\log 2 
-\frac{15}{64})\epsilon^{2}+O(\epsilon^{3})\\
\overline{\gamma^{2}}&\simeq &\frac{1}{16}g^{2}_{1}-(\frac{1}{2}\log 2
+\frac{13}{96})g^{3}_{1}\nn\\
&=& \frac{9}{64}\epsilon^{2}-(\frac{27}{16}\log 2
-\frac{99}{252})\epsilon^{3}+O(\epsilon^{4})
\eea
The expression for the correlation function (3.34) could
be given, more conveniently, in the form:
\beq
\log R^{4\Delta^{(0)}_{\sigma}}\overline{<\sigma(0)\sigma(R)>^{2}}
=\mbox{const}+2\bar{\gamma} \xi_{R}+2(\overline{\gamma^{2}}
-(\bar{\gamma})^{2})\xi^{2}_{R}+\cdots
\eeq
The term $\sim \xi_{R}$ gives correction to the effective 
scaling dimension of the operator
\beq
O_{ab}(x)=\sigma_{a}(x)\sigma_{b}(x), \,\,\,a\neq b
\eeq
\beq
\Delta''_{\sigma^{2}}=2\Delta^{(0)}_{\sigma}-\bar{\gamma}
\eeq
--comp. eqs.(3.15), (3.16) for the RS case. $\bar{\gamma}$, eq.(3.40), 
has to be compared with $\gamma_{*}$, eq.(3.12). 

The term $\sim \xi^{2}_{R}$ in (3.42) corresponds to the deviation from
the usual scaling form of the correlation function. Its coefficient 
is given by:
\beq
2(\overline{\gamma^{2}}-(\bar{\gamma})^{2})\simeq\frac{1}{8}g^{3}_{1}=\frac{27}{64}\epsilon^{3} + O(\epsilon^{4})
\eeq

In numerical simulation experiment it is more appropriate to measure
the corresponding "magnetization", instead of the correlation
function. The "magnetization" in the present case will be the
expectation value of the operator $\frac{1}{n}\sum_{a\neq b}
O_{ab}\equiv\frac{1}{n}\sum_{a\neq b} \sigma_{a}\sigma_{b}$,
i.e. $\frac{1}{n}\langle\sum_{a\neq b} O_{ab}\rangle$.  This quantity
is to be measured for a finite lattice, of size $L\times L$, at the
critical point of an infinite lattice, and then the finite size
scaling analysis is to be applied.

In the theory one obtains $<\sum_{a\neq b}O_{ab}>_{L}$, for a finite
lattice of size $L$, from the correlation function $<\sum_{a\neq
b}O_{ab}(0)\times\sum_{c\neq d}O_{cd}(R)>\propto\overline
{<\sigma(0)\sigma(R)>^{2}}$ defined on an infinite lattice, by putting
$R=L$ and taking a square root. In this way one finds, in the RSB
case:
\bea
\log\{L^{2\Delta^{(0)}_{\sigma}}
\frac{1}{n}\langle\sum_{a\neq b}O_{ab}(0)\rangle_{L}\}
&=&\log\{L^{2\Delta^{(0)}_{\sigma}}\frac{1}{n}\sum_{a\neq b}
<\sigma_{a}(0)>_{L}<\sigma_{b}(0)>_{L}\}\nn\\
&=&\mbox{const}_{1} +\bar{\gamma}\xi_{L}+(\overline{\gamma^{2}}-
\overline{\gamma^{2}})\xi^{2}_{L}+\cdots
\eea
The operator $O_{ab}(x)=\sigma_{a}(x)\sigma_{b}(x)$, in the theory, 
is a product of local spins for different replicas. In numerical simulations 
$O_{ab}(x)$ will be a local product of spins for two copies of the same 
random lattice, simulated with different initial conditions for spins. 
Two identical disordered lattices, different starting conditions, they
realize in numerical experiment two replicas of the theory. 
We shall have first
\beq
<O_{ab}(x)>_{L}=<\sigma_{a}(x)>_{L}<\sigma_{b}(x)>_{L}
\eeq
-- for a given disorder, and then the average over the disorder, 
is to be taken. According to the replica theory of spin-glasses [6]
in the situation when the replica symmetry is broken, the space of states
is divided into many "valleys" corresponding to different ground states,
and the summation over replica indices in the thermally averaged quantities
corresponds to the summation over all these "valleys".
Since different initial conditions, in general, correspond to
different ground states, the summation over various samples (for averaging
over the disorder) with different initial conditions must correspond
to the summation over the indices $a,b$ in the quantity (3.47).

In the RS case, one obtains from eqs.(3.14), (3.15):
\bea
\log\{L^{2\Delta^{(0)}_{\sigma}}\sum_{a\neq b}<O_{ab}(0)>_{L}\}
&=&\log \{L^{2\Delta^{(0)}_{\sigma}}
\sum_{a\neq b}<\sigma_{a}(0)>_{L}<\sigma_{b}(0)>_{L}\}\nn\\
&=& \mbox{const}_{2}+\gamma_{*}\xi_{L}
\eea

\section{Analyses of the results. Distributions.}

For the purpose of numerical simulation tests we shall give here the
numbers for the coefficients, for the case of the 3 - component Potts
model, $\epsilon=\frac{2}{15}$. One obtains:
\beq
\gamma_{\ast}\simeq\frac{3}{8}\epsilon-(\frac{9}{4}\log 2-\frac{33}{64})\epsilon^{2}
\simeq 0.050-0.019=0.031
\eeq
-- for the RS case, eqs.(3.12), (3.48).
\beq
\bar{\gamma}\simeq\frac{3}{8}\epsilon-(\frac{9}{4}\log 2-\frac{15}{64})\epsilon^{2}
\simeq 0,050-0,024=0,026
\eeq
\beq
\overline{\gamma^{2}}-(\bar{\gamma})^{2}\simeq\frac{27}{128}\epsilon^{3}=0,0005
\eeq
-- for the RSB case, eqs.(3.40), (3.45), (3.46).

Characteristic numbers to look  at 
are the values of the products $\bar{\gamma}\xi_{L}$, 
$(\overline{\gamma^{2}}-(\bar{\gamma})^{2})\xi_{L}^{2}$ 
in eq.(3.46), for $L$ maximal in numerical simulations, $L=10^{3}$. One
gets,
\beq
\bar{\gamma}\xi_{L}=0.18
\eeq
\beq
(\overline{\gamma^{2}}-(\bar{\gamma})^{2})\xi^{2}_{L}=0.02
\eeq
This gives an estimate of $\sim10\%$ on the deviation from scaling, in
case of the 3-component Potts model. This is only an estimate as we
could not know in advance the accuracy of the $\epsilon $ -expansion
calculation. In particular, the scaling violation term (4.5) might
still be smaller. For this reason we have also done simulations for
the 4-component Potts model. The results will be presented in the next
Section. In the theory, the $\epsilon$ - expansion values for the
coefficients do not make sense for the 4-component model
$(\epsilon=\frac{1}{3})$. In particular, the second term in the
eq.(3.40) for $\bar{\gamma}$ becomes bigger than the first one, and
$\bar{\gamma}$ becomes negative. On the other hand the model itself
should evolve continuously with the number of components $q$, up to 4
and further. This is because the phase transition of the Potts model
with random bonds remains second order for $q>4$ [9,10,11]. Then we
would expect that the effect of the deviation from scaling, if
present, should become more pronounced as $q$ is increased. So it
should be easier observable for the 4-component model, compared to the
3-component one. Going from $q=3$ to $q=4$ we go out of the
perturbative region, where the $\epsilon$ expansion is valid, but the
qualitative effect of scaling violation should increase, if the model
is at the RSB fixed point.

In addition to the above results on the correlation function squared
and the associated magnetization, one could also look at the
corresponding distributions. This implies that after the calculation
of the thermodynamic expectation values of $<~\sigma_{a}(x)~>$ and
$<\sigma_{b}(x)>$, for two identical lattices, different starting
conditions, one changes disorder and performs the thermodynamic
measurement again, and gets another values of
$<~\sigma_{a}(x)>,\,\,<\sigma_{b}(x)>.$ Having done these measurements
many times, one constructs the distribution for the values of products
of local magnetizations
\beq
<O_{ab}(x)>_{L}=<\sigma_{a}(x)>_{L}<\sigma_{b}(x)>_{L}
\eeq
This is instead of summing up the values for products and calculating
in this way the average over the disorder. [We observe that at the
stage of calculating either the distribution or the average over the
disorder of the products (4.6) one could use also the values of local
products for different points $x$ on the lattice, if they have been
measured. But we stress again that local products have to be taken
first, summation over $x$ second.]

The distribution obtained in this way, the ``overlap function'' of
local magnetizations, could be obtained in the theory from the RG
result for the amplitude
\beq
Z(\zeta_{L},t)\sim e^{\gamma(t)\xi_{L}}
\eeq
It is more convenient to study the $\log $ of this function:
\beq
Q(t)=\log Z(\xi_{L},t) = \mbox{const}+\gamma(t)\xi_{L}\equiv Q_{0}+\gamma(t)\xi_{L}
\eeq
which corresponds to 
\beq
Q_{ab}=\log\{(L)^{2\Delta^{(0)}_{\sigma}}<\sigma_{a}(x)>_{L}<\sigma_{b}(x)>_{L}\}
\eeq
We shall define
\beq
q(t)=Q(t)-Q_{0}
\eeq
In analogy with the theory of spin-glasses, to obtain the distribution 
of values of $Q_{ab}$
one has to define the inverse function of $Q(t)$, or of $q(t)$, 
and calculate its derivative.
One gets:
\bea
q(t)&=&\gamma(t)\xi_{L}\\
\gamma(t)&=&\cases{\frac{t}{12},\,\,\, 0<t<t_{1}(=3g_{1})\cr
\gamma_{1},\,\,\, t_{1}<t<1\cr}
\eea
\bea
\gamma_{1}=a_{1}g_{1}-a_{2}g^{2}_{1}=
\frac{1}{4}g_{1}-(\log 2+\frac{1}{48})g^{2}_{1}\nn\\
=\frac{3}{8}\epsilon-(\frac{9}{4}\log 2-\frac{69}{64})\epsilon^{2}
\eea
We have used equations (3.29)--(3.31) 
for $\gamma(t)$ and we have been keeping the accuracy
of our calculations by dropping extra terms. In particular, 
in the interval $0<t<t_{1}$ we
can keep linear terms only. 
This is because the interval itself is $\sim\epsilon $ and under
integration the linear terms become quadratic, $\sim\epsilon^{2}$. One has:
\beq
q(t)=\cases{\frac{t}{12}\xi_{L},\,\,\,0<t<t_{1}\cr q_{1},\,\,\,t_{1}<t<1\cr}
\eeq
\beq
q_{1}=\gamma_{1}\xi_{L}
\eeq
In the interval $0<q<q_{1}$ the inverse function is given by:
\beq
t(q)=\frac{12}{\xi_{L}}q,\,\,\,0<q<q_{1}
\eeq
Its derivative:
\beq
\frac{dt(q)}{dq}=\frac{12}{\xi_{L}},\,\,\,O\leq q<q_{1}
\eeq
Finally one gets the following distribution function:
\beq
N(q)=\frac{dt}{dq}=\cases{\frac{12}{\xi_{L}}+(1-3g_{1})\delta(q-q_{1}),\,\,\,
0<q\leq q_{1}\cr 0,\,\,\,q>q_{1}\cr}
\eeq

Numerical measurement of the distribution $N(q)$ might be complicated.
For the product of magnetizations on finite lattices of size $L$,
eq.(4.9), the distribution $N(q)$, will actually be rounded by finite
size effects. To have it more distinct one have to look for the limit
of big $L$, but then the extra structure in $N(q)$, for $0<q<q_{1}$,
would become lower and lower, being of the hight 
$\sim\frac{1}{\xi_{L}}$.

The way out could be to look at the distribution 
of a products of correlation functions themselves:
\beq
Q_{ab}^{(2)}(R)=\log\{(R)^{4\Delta^{(0)}_{\sigma}}<\sigma_{a}(0)
\sigma_{a}(R)>_{L}
<\sigma_{b}(0)\sigma_{b}(R)>_{L}\}
\eeq
The index (2) of $Q_{ab}^{(2)}$ is meant to tell that we are looking at 
the overlap function of a two-point object. The above considered $Q_{ab}$ 
for magnetizations could have been noted $Q_{ab}^{(1)}$.

In the theory, the corresponding $Q^{(2)}(R,t)$ shall be given 
by $Z^{2}(\xi_{R},t)$:
\beq
Q^{(2)}(R,t)=\log Z^{2}(\xi_{R},t)
\eeq
$\xi_{R}=\log R$, and it is assumed that $1\ll R\ll L,\,\,L$ 
being the size of the lattice.
$R$ has to be big enough so that the continuum limit theory 
applies and the fixed point is reached. In an analogous way 
one finds in this case
\beq
N^{(2)}(q)=\frac{dt}{dq}=\cases{\frac{6}{\xi_{R}}+(1-3g_{1})\delta(q
-q_{1}),\,\,\,0<q\leq q_{1}\cr 0,\,\,\,q>q_{1}\cr}
\eeq
Now, as we increase $L$, the profile of $N^{(2)}(q)$ will be sharpened 
while the hight of the extra structure $\sim\frac{1}{\xi_{R}}$ 
remain unchanged.

\section{Simulations.}
In this section, we are going to present some results of numerical
simulations that we performed in order to check the validity of our
results. In particular, we want to try to find a way to choose between the
two possibles scenarios, replica breaking or replica symmetry. The easiest
thing that we can simulate, as explained earlier, is the scaling dimension
of the square magnetization. Thus we have performed the following
simulations: On a square lattice of size $L\times L$, we simulate two
configurations ($\sigma^a$ and  $\sigma^b$) of the $q$-state Potts model
with the {\it same} disorder, 
but with different initial conditions. Then we compute the product of the
magnetization 
\begin{equation}
Q_{ab} = {1\over L^2}\sum_{i=1,L^2}<\sigma^a_i> <\sigma^b_i > 
\end{equation} 
Here $<\sigma^a_i>$ means the thermal average of the local magnetization 
\begin{equation}
\sigma^a_i = \vec{\sigma^a_i}\cdot \vec{m^a}
\end{equation} 
and $\vec{m^a}$ is the total magnetization 
\begin{equation}
\vec{m^a}={1\over L^2} \sum_{i=1,L^2} \vec{\sigma^a_i}
\end{equation}
In practice, as we have checked numerically, it turns out that this
quantity is the same as
\begin{equation}
Q_{ab} = {1\over L^2}\sum_{i=1,L^2}<\sigma^a_i \sigma^b_i > 
\end{equation} 
The Hamiltonian of the simulated model is given by
\begin{equation}
H=-\sum_{\{i,j\}}
J_{ij}(\delta_{\sigma^a_i,\sigma^a_j}+\delta_{\sigma^b_i,\sigma^b_j}) 
\end{equation}
where the coupling constant between nearest neighbor spins takes the value
\[ J_{ij}=\left\{\begin{array}{ll}
                J_0 & \mbox{with probability $p$}\\
                J_1 & \mbox{with probability $1-p$}
               \end{array}
\right. \]
Measurements were performed on a square lattice with helical boundary
conditions. Without any lost of 
generality, we can consider the case where $p={1\over 2}$. Then the model
is self-dual and thus the critical temperature is exactly known. It
is given by the solution of the equation \cite{kd}
\begin{equation}
{1-e^{-\beta J_0} \over 1+(q-1)e^{-\beta J_0}}  =  e^{-\beta J_1}.
\end{equation}
The disorder that we choose to simulate is $J_0=1, J_1={1\over 10}$. 
This disorder is in fact quit strong in order to avoid problems of
cross-over \cite{mp}.
Monte Carlo data were obtained by using the well known Wolff cluster
algorithm \cite{wolf}. Due to the strong disorder
that we considered, we needed to have large statistics over the number of
configurations of disorder. Simulations were performed for lattice with
size ranging from $L=10$ to $L=1000$. The number of
configurations of disorder were $20 000$ for $L=20-200$, $6 000$ for
$L=500$ and $1 000$ for $L=1000$. For each of these configuration of
disorder, measurements were taken over $t_1$ updates, after $t_0$
updates for thermalisation. The statistical error $\delta A$ of 
a quantity $A$ has two contributions, one from the thermal fluctuation, 
with a variance $\sigma_T$, and one from the disorder fluctuation, with a 
variance $\sigma_N$. Thus the statistical error is given by
\beq
(\delta A)^2 = {\sigma^2_N \over N} + {\sigma^2_T \over N t_1/\tau}
\eeq
where $N$ is the number of configurations of disorder and $\tau$ is 
the autocorrelation time. For the quantity that we measured, it turns out 
that the two variance are near equal and then
\beq
(\delta A)^2 \simeq {\sigma^2_N \over N}(1 + {\tau \over  t_1 })
\eeq
Thus we just need to choose $t_1$ such that $t_1 >> \tau$ and we can
ignore the thermal fluctuations. Then the first step is to measure the
autocorrelation time. For the $3$-state Potts model, we measured $\tau
\simeq 3$ for L=10 up to $\tau \simeq 25$ for $L=1000$. Thus by
choosing $t_0=t_1=1000$, we can safely ignore the thermal
fluctuations.  For the $4$-state Potts model, we measured $\tau \simeq
5$ for L=10 up to $\tau \simeq 50$ for $L=1000$.  Again, we choosed
the parameters $t_0=t_1=1000$ except for $L=1000$ for which we took
$t_0=t_1=2000$ and this in order to be sure to have thermalized data.

Results of these measurements for the $3$-state Potts model are 
displayed in Fig. 1. 
%
%
%
%
\begin{figure}
\epsfxsize=400pt\epsffile{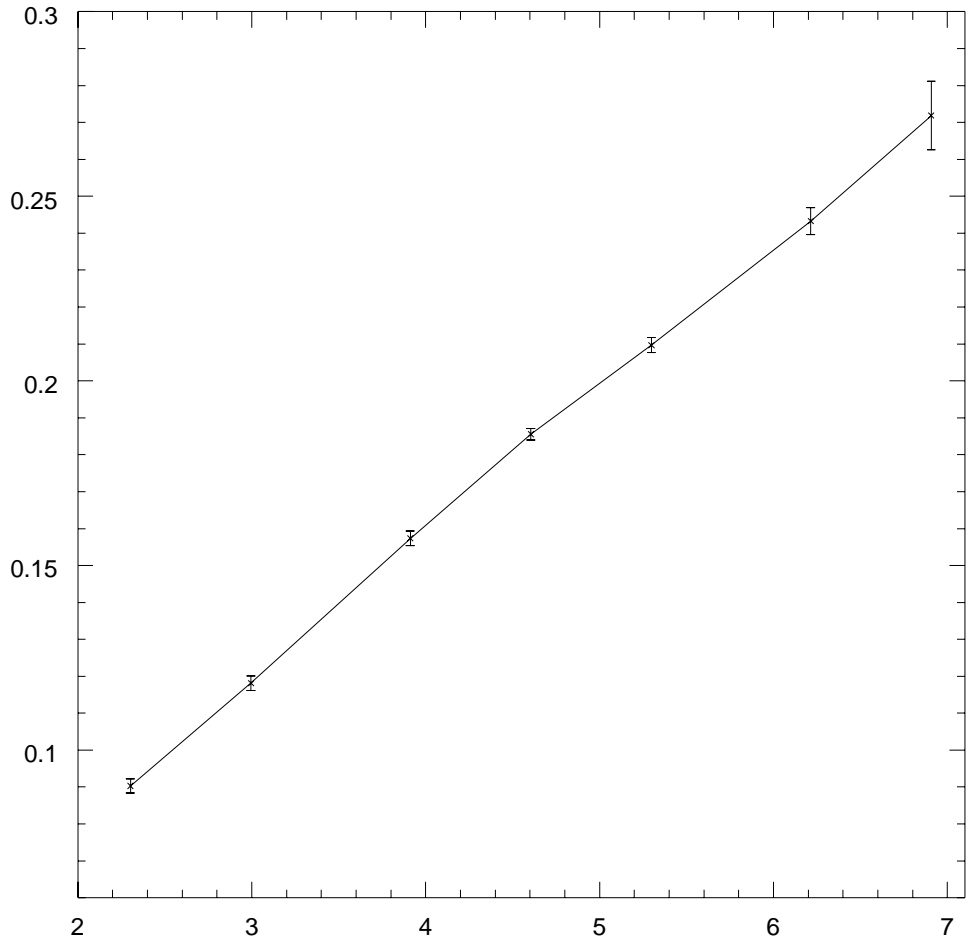}
  \caption{
  Plot of $\ln(L^{2\Delta_\sigma} \overline{Q_{ab}})$ vs. $\ln(L)$ 
  for the 3-state Potts model.}
\label{Fig3p}
\end{figure}

In this figure, we plot $\ln(L^{2\Delta_\sigma} \overline{Q_{ab}})$
versus $\ln(L)$. Here, $\overline{(\cdots)}$ means the average over
the disorder.  As explained in section 3, we expected two possible
behaviors for this quantity, either $\ln(L^{2\Delta_\sigma}
\overline{Q_{ab}}) \simeq \mbox{const}_{1} +\gamma_{*}\ln(L)
+\cdots$ in the RS scenario (see eq. (3.48)) or $\ln(L^{2\Delta_\sigma}
\overline{Q_{ab}}) \simeq \mbox{const}_{1} +\bar{\gamma}\ln(L)
+(\overline{\gamma^{2}}- \overline{\gamma^{2}})\ln^2(L)
+\cdots$ in the RSB scenario (see eq. (3.46)). According to the RS scenario, 
we would expect a scaling behavior {\it ie.} a linear (Log-Log) plot, 
which is in very good agreement 
with what we obtain. Moreover, we estimate that, for the RSB scenario, 
the deviation from such a linear behavior due to the $\ln^2(L)$ term should 
be of
order $10\%$ at $L=1000$. We do not see such a deviation. Thus our 
numerical simulation of the $3$-state Potts model does clearly favor 
the replica symmetry solution over the replica symmetry breaking one.

Performing a fit of the plot
\begin{equation}
\label{fp}
L^{2\Delta_\sigma}\overline{Q_{ab}} \simeq L^{\gamma_{*}}
\end{equation}
we obtain a value of $\gamma_{*}=0.04\pm 0.002$, which is reasonably close to 
the predicted value $\gamma_{*}=0.031$ (see eq. (4.1)).

Fig. 2. corresponds to the same plot but for the $4$-state Potts model. 
As explained in section 4, we do not expect that the $\epsilon$-expansion 
is still valid for $q=4$, but we would still expect to see a deviation 
from scaling due to the RSB and we expect that this deviation is more 
pronounced as we increase $q$. But again, we do not see a deviation 
from a scaling behavior, thus again not seeing any evidence in favor 
of the replica symmetry breaking scenario.
%
%
%
%
\begin{figure}
\epsfxsize=400pt\epsffile{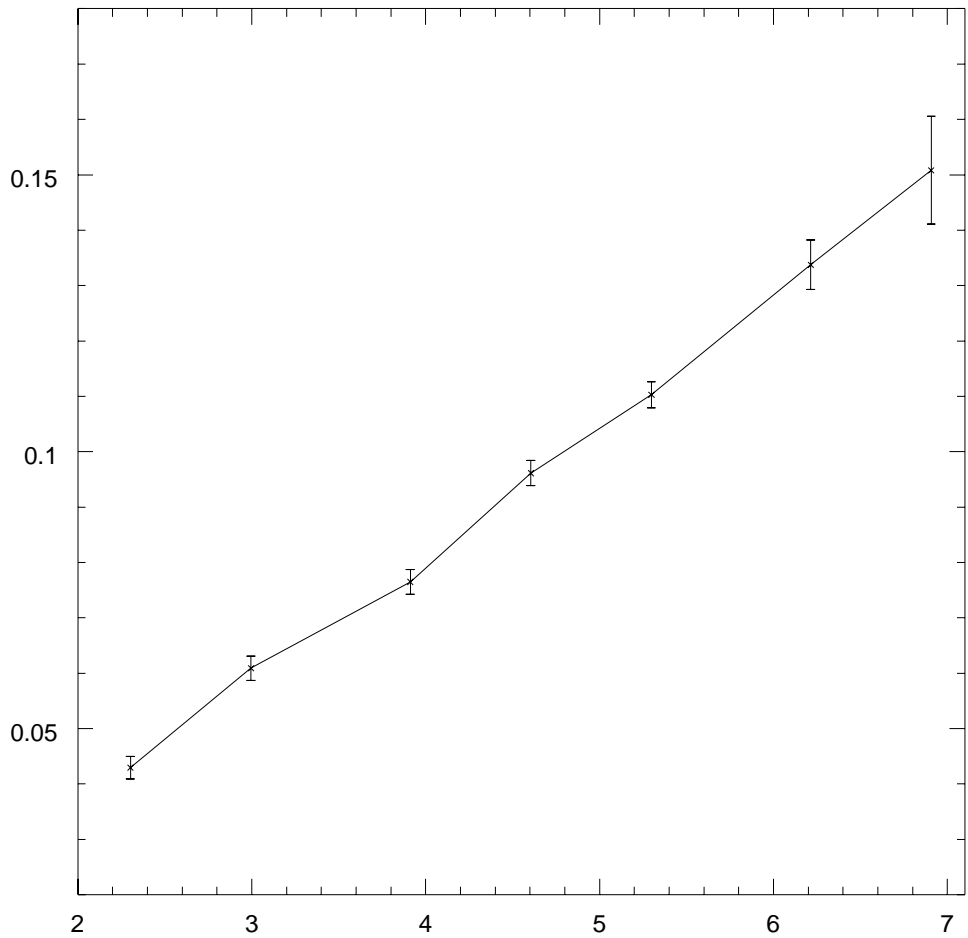}
  \caption{
  Plot of $\ln(L^{2\Delta_\sigma} \overline{Q_{ab}})$ vs. $\ln(L)$ 
  for the 4-state Potts model.}
\label{Fig4p}
\end{figure}
Performing a fit of the plot with eq.(\ref{fp}), we do obtain here $\gamma_{*}=0.023 \pm 0.002$.

\section{Conclusions and Discussions.}

We consider that the results of numerical simulations presented in the
preceding Section support the RS fixed point critical behavior.
Neither for the 3-component nor for the 4-component models could we
detect the deviation from scaling, characteristic of the RSB fixed
point. In case of the 3-component model, for which the
$\epsilon$-expansion could be expected to be reasonably well defined,
the value of the slope of the curve for the scaling function (3.48)
agrees sufficiently well with the value of $\gamma_{\ast}$ in
eq.(4.1).

Still we would like to make a remark on possible physical significance
of RSB fixed point, so that it would not appear totally formal.

The coupling constant $g_{ab}$ in (1.13) is proportional to the
overlap function of local energies, in a similar way as $Z_{ab}$ is
proportional to the overlap function of spins. Having $g_{ab}$
different for different $a,b$, as it is the case for the RSB fixed
point, could be attributed to a multiplicity of ground states in the
statistical model.  One starts with different initial conditions for
spins (one breaks initially equivalence of replicas in this way) and
the model can end in different ground states.  In turn, multiplicity
of ground states could be interpreted as a kind of localization
phenomenon, in the configurational space of spins. Unlike in
spin-glasses, in the present model with relatively weak disorder the
localization would be due to fluctuations, which make disorder
important at large distances. It is this spontaneous phenomenon that
we had in mind behind the notion of RSB in the critical model with
disorder.

We would also like to remark that one of the possibilities, why the
RSB phenomena have not been detected in the particular models
considered, could be due to the so called "marginal stability" of the
RSB fixed point: it is well known that in the linear order stability
analysis of the RSB fixed point (3.24) there exists the so called
"zero-mode" with the zero eigenvalue. The detailed second order
calculations of the stability of this fixed point shows that to enter
the critical regime defined by the RSB fixed point (3.24), in general
one needs to reach exponentially large spatial scales [16], which
could be well beyond the lattice sizes of the present numerical tests.

\vskip 2cm
\noindent{\large\bf Acknowledgments}

This research was supported in part by National Science Foundations
under Grant No. PHY 94-07194. Vl.D. and M.P. thank the Institute for
Theoretical Physics at Santa Barbara, where part of this work was
done, for its hospitality.  Vik.D would like to acknowledge partial
support by Russian Fund for Fundamental Research, Grants No
96-02-18985 and No 96-15-96920.  M. P. would like to acknowledge
partial support by CAPES (Brazil) and thanks the Physics Department
of the UFES, and in particular J.~Fabris, for its hospitality while
this work was being completed.


\appendix
\section{Appendix}
\subsection{The integral of $D^{(2)}_{1},$ eq.(2.17).}

\beq
I_{1}=\int d^{2}y\int d^{2}y'\frac{1}{|y|^{\Delta_{\varepsilon}}}
\frac{1}{|y'|^{\Delta_{\varepsilon}}}\frac{1}{|y-y'|^{2\Delta_{\varepsilon}}}
\eeq
Here $\Delta_{\epsilon}=1-\frac{3}{2}\epsilon $. We change the 
variable $y':\,\,y'=yt$. This gives:
\bea
I_{1}&=& \int d^{2}y|y|^{2-4\Delta_{\varepsilon}}\int 
d^{2}t|t|^{-\Delta_{\varepsilon}}|1-t|^{-2\Delta_{\varepsilon}}\nn\\
&=&\int_{1<|y|<a} d^{2}y|y|^{-2+6\epsilon}\int d^{2}t|t|^{2\alpha}
|1-t|^{2\beta}\nn\\
&=&2\pi\frac{1}{6\epsilon}(a^{6\epsilon}-1)\pi\frac{\Gamma(1
+\alpha)\Gamma(1+\beta)\Gamma(-1-\alpha-\beta)}{\Gamma(-\alpha)
\Gamma(-\beta)\Gamma(2+\alpha+\beta)}
\eea
Here $\alpha=-\frac{\Delta_{\varepsilon}}{2}=-\frac{1}{2}+\frac{3}{4}\epsilon,\,\,\,\beta
=-\Delta_{\varepsilon}=-1+\frac{3}{2}\epsilon$, and we have used the known result:
\beq
\int d^{2}t|t|^{2\alpha}|1-t|^{2\beta}
=\pi\frac{\Gamma(1+\alpha)\Gamma(1+\beta)\Gamma(-1-\alpha
-\beta)}{\Gamma(-\alpha)\Gamma(-\beta)\Gamma(2+\alpha+\beta)}
\eeq
For its derivation see e.g.[12]. Putting the values of $\alpha,\beta $ and 
expanding in $\epsilon$
one obtains:
\bea
\frac{\Gamma(1+\alpha)\Gamma(1+\beta)\Gamma(-1-\alpha-\beta)}{\Gamma(-\alpha)
\Gamma(-\beta)\Gamma(2+\alpha+\beta)}&\approx &
\frac{2}{3\epsilon}(1+3\epsilon(\psi(1)-\psi(\frac{1}{2})))\nn\\
&=& \frac{2}{3\epsilon} (1+\epsilon 6\log 2)
\eea
Finally one gets:
\beq
I_{1}=2\pi^{2}(1+\epsilon K)\frac{1}{9\epsilon^{2}}(a^{6\epsilon}-1)
\eeq
$K=6\log 2.$

\subsection{The integral of $D^{(2)}_{2}$, eq.(2.19).}

\beq
I_{2}=\int d^{2}y\int d^{2}y'<\sigma(0)\varepsilon(y)\varepsilon (y')\sigma(\infty)><\varepsilon(y)
\varepsilon(y')>
\eeq
This integral has already appeared in the calculations in the paper [2]. It is finite, the limit
of $\epsilon\rightarrow 0$. This could also be seen without calculations, by using the operator
algebra. In fact, when $y\rightarrow 0$ one has:
\beq
\sigma(0)\varepsilon(y)=\frac{D}{|y|^{\Delta_{\varepsilon}}}\sigma(0)+\cdots
\eeq
As $\Delta_{\varepsilon}\approx 1$, the integration over $y$ around $0$ is finite.

When $y,y'\rightarrow 0,\,\,y'\gg y$, one has:
\beq
\sigma(0)\varepsilon(y)\varepsilon(y')
=\frac{D^{2}}{|y|^{\Delta_{\varepsilon}}|y'|^{\Delta_{\varepsilon}}}
\sigma(0)+\cdots
\eeq
Again, the integration around $0$ is finite.

Finally, for the configuration of $y'\rightarrow y$ one obtains:
\beq
\varepsilon(y)\varepsilon(y')=\frac{1}{|y-y'|^{2\Delta_{\varepsilon}}}(1 
+\cdots+|y-y'|^{4}T(y)\bar{T} (\bar{y})+\cdots)+\cdots
\eeq
$T(y), \bar{T}(\bar{y})$ are components of the energy-momentum
operator. In the OPE (A.9) the term $\sim T\bar{T}$ is the first after
1 which contributes when integrated in (A.6). Some extra terms, which
vanish under integration, like $(y-y')^{2}T(y)$ or
$(\bar{y}-\bar{y'})^{2}\bar{T}(\bar{y})$ have been dropped. Putting
$\Delta_{\varepsilon}\approx 1$, one gets
\bea
&&<\sigma(0)\varepsilon(y)\varepsilon(y')\sigma(\infty)> 
<\varepsilon(y)\varepsilon(y')>\\
&&\qquad =<\sigma(0)\sigma(\infty)>\frac{1}{|y-y'|^{4}}(1+
|y-y'|^{4}<\sigma(0)T(y)\bar{T}(y)\sigma
(\infty)>+\cdots)+\cdots \nn
\eea
The first term, $\sim 1/|y-y'|^{4}$, leads to quadratic divergence in
the integral (A.6). It is automatically subtracted in the analytic
$\epsilon$-expansion calculation. It corresponds, in fact, to the
constant shift renormalization of the action of the theory which is
irrelevant. The next term is finite, when $y'\rightarrow y$. So the
integral (A.6) will be finite.

\subsection{The integral of $D^{(2)}_{3}$, eq.(2.21).}

\beq
I_{3}=\int d^{2}y'\int d^{2}y(<\sigma(0)\varepsilon(y)
\varepsilon(y')\sigma(\infty)>)^{2}
\eeq

The variable $y'$ could be scaled out by using invariance of
correlation functions w.r.t. global dilatation. In general, when a set
of operators $O_{1},O_{2},\cdots,O_{n}$ is projected on the operator
$O_{n+1}$, placed at infinity, one has:
\bea
&&<O_{1}(x_{1})O_{2}(x_{2})\cdots O_{n}(x_{n})O_{n+1}(\infty)>\nn\\
&&=\lambda^{\Delta_{1}+\Delta_{2}\cdots +\Delta_{n}
-\Delta_{n+1}}<O_{1}(\lambda x_{1})O_{2}(\lambda x_{2})\cdots O_{n}
(\lambda x_{n})O_{n+1}(\infty)>
\eea
where $\lambda $ is a dilatation parameter. For
$<\sigma(0)\varepsilon(y)\varepsilon(y')\sigma (\infty)>$ this gives:
\beq
<\sigma(0)\varepsilon(y)\varepsilon(y')\sigma(\infty)>=
\lambda^{2\Delta_{\varepsilon}}<\sigma(0)
\varepsilon(\lambda y)\varepsilon(\lambda y')\sigma(\infty)>
\eeq
Putting $\lambda=1/y'$, one obtains:
\beq
<\sigma(0)\varepsilon(y)\varepsilon(y')\sigma(\infty)>=|y'|^{-2\Delta_{\varepsilon}}
<\sigma(0)\varepsilon(\frac{y}{y'})\varepsilon(1)\sigma(\infty)>
\eeq
Using this relation in the integral (A.11) and changing the variable 
$y,\,\,y=y'\tilde{y},$ one obtains:
\beq
I_{3}=\int d^{2}y'|y'|^{2-4\Delta_{\varepsilon}}\int d^{2}\tilde{y}
(<\sigma(0)\varepsilon(\tilde{y}) \varepsilon(1) \sigma(\infty)>)^{2}
\eeq
Putting $\Delta_{\varepsilon}=1-\frac{3}{2}\epsilon $ and integration $y'$ 
between the cut-offs 
$1,a$, one gets:
\beq
I_{3}=\int_{1<|y'|<a} d^{2}y'|y|^{-2+6\epsilon}\tilde{I}_{3}=
2\pi\frac{1}{6\epsilon}(a^{6\epsilon}-1)\tilde{I}_{3}
\eeq
\beq
\tilde{I}_{3}=\int d^{2}y(<\sigma(0)\varepsilon(y)\varepsilon(1)\sigma(\infty)>)^{2}
\eeq
In the Coulomb gas representation the correlation function 
$<\sigma\varepsilon\varepsilon\sigma>$ could be presented in the following form:
\bea
&&<\sigma(0)\varepsilon(y)\varepsilon(1)\sigma(\infty)>\nn\\
&&=N \int d^{2}u<V_{\bar{\alpha}_{\sigma}}(0)
V_{\alpha_{\varepsilon}}(y)
V_{\alpha_{\varepsilon}}(1)V_{\alpha_{+}}(u)V_{\alpha_{\sigma}}(\infty)>\nn\\
&&=N\int d^{2}u|y|^{4
\bar{\alpha}_{\sigma}\alpha_{\varepsilon}}|y-1|^{4\alpha^{2}_{\varepsilon}}
|u|^{4\bar{\alpha}_{\sigma}\alpha_{+}}|y-u|^{4\alpha_{\varepsilon}
\alpha_{+}}|1-u|^{4\alpha_{\varepsilon}\alpha_{+}}
\eea
Here $N$ is the normalization constant. It can be fixed by using the 
operator algebra:
\bea
r &\rightarrow & 0,\quad \sigma(0)\sigma(r)
=\frac{1}{|r|^{2\Delta_{\sigma}}}+\cdots\\
r &\rightarrow & 0,\quad V_{\bar{\alpha}_{\sigma}}(0)V_{\alpha_{\sigma}}(r)
=\frac{1}{|r|^{2\Delta_{\sigma}}} V_{2\alpha_{0}}(0)+\cdots
\eea
For the correlation function
\bea
&&<\sigma(0)\sigma(r)\varepsilon(y)\varepsilon(y')>\nn\\
&&=N\int d^{2}u<V_{\bar{\alpha}_{\sigma}}(0)
V_{\alpha_{\sigma}}(r)V_{\alpha_{\varepsilon}}(y)
V_{\alpha_{\varepsilon}}(y')V_{\alpha_{+}}(u)>
\eea
This gives, in the limit of $r\rightarrow 0$:
\bea
\frac{1}{|r|^{2\Delta_{\sigma}}}<\varepsilon(y)\varepsilon(y')>
&=&N\frac{1}{|r|^{2\Delta_{\sigma}}}
\int d^{2}u<V_{2\alpha_{0}}(0)V_{\alpha_{\varepsilon}}(y)V_{\alpha_{\epsilon}}
(y')V_{\alpha_{+}}(u)>\\
\frac{1}{|y-y'|^{2\Delta_{\varepsilon}}}
&=& N|y|^{8\alpha_{0}\alpha_{\varepsilon}}
|y'|^{8\alpha_{0}\alpha_{\varepsilon}}|y-y'|^{4\alpha^{2}_{\varepsilon}}\nn\\
&&\times\int d^{2}u|u|^{8\alpha_{0}\alpha_{+}}|u-y|^{4\alpha_{\varepsilon}
\alpha_{+}}|u-y'|^{4\alpha_{\varepsilon}\alpha_{+}}
\eea
Changing the variable $u$ in the integral:
\beq
u=\frac{yy'}{y-y'}\times\frac{1}{\tilde{u}+\frac{y'}{y-y'}}
\eeq
one obtains, after simple algebra:
\beq
\frac{1}{|y-y'|^{2\Delta_{\varepsilon}}}=N\frac{1}{|y-y'|^{2
\Delta_{\varepsilon}}}\int d^{2}
\tilde{u}|\tilde{u}|^{-2\alpha^{2}_{+}}|\tilde{u}-1|^{-2\alpha^{2}_{+}}
\eeq
Here $2\Delta_{\varepsilon}=4\Delta_{12}=4\alpha^{2}_{12}-8\alpha_{12}\alpha_{0}$.
Using $\alpha_{12}=-\alpha_{+}/2,\,\,2\alpha_{0} 
=\alpha_{+}+\alpha_{-},\,\,\alpha_{+}\alpha_{-}=1$, one gets 
also $2\Delta_{\varepsilon}=-2+3\alpha^{2}_{+}$. From (A.25):
\beq
N=(\int d^{2}u|u|^{-2\alpha^{2}_{+}}|u-1|^{-2\alpha^{2}_{+}})^{-1}
\eeq
Here $\alpha^{2}_{+}=\frac{4}{3}-\epsilon$.
One obtains:
\beq
N=\pi^{-1}\frac{\Gamma^{2}(\alpha^{2}_{+})\Gamma(2-2\alpha^{2}_{+})}
{\Gamma^{2}(1-\alpha^{2}_{+})
\Gamma(-1+2\alpha^{2}_{+})}=\pi^{-1}\frac{\Gamma^{2}(\frac{4}{3}-\epsilon)
\Gamma(-\frac{2}{3}+2\epsilon)}
{\Gamma^{2}(-\frac{1}{3}+\epsilon)\Gamma(\frac{5}{3}-2\epsilon)}
\eeq
We have used again the integral (A.3). In our further calculations the
normalization constant N will multiply expressions with singularities
$\sim 1/\epsilon$, but not $1/\epsilon^{2}$. This implies that we
don't need to keep the $\epsilon $ correction of N. For $\epsilon=0$,
and using the usual properties of $\Gamma$-functions, one finds from
(A.27):
\bea
N &=& -\frac{2\Gamma^{3}(-\frac{2}{3})}{9\pi\Gamma^{3}(-\frac{1}{3})}
=-\frac{2}{9\sin(\pi(-\frac{2}{3}))}
\frac{\Gamma^{2}(-\frac{2}{3})}{\Gamma(\frac{5}{3})\Gamma^{3}(-\frac{1}{3})}
\nn\\
&=& -\frac{2}{\sqrt{3}}\frac{\Gamma
^{2}(-\frac{2}{3})}{\Gamma^{4}(-\frac{1}{3})}
\eea
or
\beq
N=-\frac{2}{\sqrt{3}}\gamma^{-2},\,\,\,\gamma=\frac{\Gamma^{2}(-\frac{1}{3})}
{\Gamma(-\frac{2}{3})}
\eeq
Returning to the integral $\tilde{I}_{3}$ in (A.17) and using the
expression (A.18) for the correlation function
$<\sigma\varepsilon\varepsilon\sigma>$, one gets:
\beq
\tilde{I}_{3}=N^{2}I
\eeq
\bea
I&=&\int d^{2}y \int d^{2}u_{1}\int d^{2}u_{2}|y|^{4\bar{\alpha}_{\sigma}
\alpha_{\varepsilon}}|y-1|^{4\alpha^{2}_{\varepsilon}}
|u_{1}|^{4\bar{\alpha}_{\sigma}\alpha_{+}}|u_{1}-1|^{4
\alpha_{\varepsilon}\alpha_{+}}|u_{1}-y|^{4
\alpha_{\varepsilon}\alpha_{+}}\nn\\
&&\qquad \times|u_{2}|^{4\bar{\alpha}_{\sigma}
\alpha_{+}}|u_{2}-1|^{4\alpha_{\varepsilon}\alpha_{+}}
|u_{2}-y|^{4\alpha_{\varepsilon}\alpha_{+}}
\eea
The integral $I$ is calculated in the Appendix B, with the result:
\beq
I=-\frac{\pi}{16}\gamma^{4}+O(\epsilon)
\eeq
$\gamma $ is defined in (A.29). For $\tilde{I}_{3}$ one gets:
\beq
\tilde{I}_{3}=N^{2}I=-\frac{\pi}{12}+O(\epsilon)
\eeq
and then for the integral $I_{3}$, (A.11), one obtains from (A.16):
\beq
I_{3}=\cdots -\frac{\pi^{2}}{6}\frac{1}{6\epsilon}(a^{6\epsilon}-1)+\cdots
\eeq

This expression is in fact incomplete. It has to be corrected. There
is an extra term in it, $\sim 1/\epsilon^{2}$, which is missed in the
analytic technique of calculating the integral $I$ in the Appendix
B. It could be recovered by using the original expression for the
integral $I_{3}$ in the eq.(A.11) and the operator
algebra. Configuration which is responsible for the leading
singularity in $I_{3}$ is either
\beq
1\ll|y|\ll|y'|,\,\,\,y,y'\rightarrow 0
\eeq
or
\beq
1\ll|y'|\ll|y|,\,\,\,y,y'\rightarrow 0
\eeq
For (A.35) one has, by operator algebra:
\bea
\sigma(0)\varepsilon(y)\varepsilon(y')&=&\frac{D}{|y|^{\Delta_{\varepsilon}}}
\sigma(0)\varepsilon(y')+\cdots\nn\\
&=&\frac{D}{|y|^{\Delta_{\varepsilon}}}\frac{D}{|y'|^{\Delta_{\varepsilon}}}
\sigma(0)+\cdots
\eea
$D=D^{\sigma}_{\sigma\varepsilon}$ is the OA coefficient. For the integral 
$I_{3}$, this gives:
\bea
&&\int d^{2}y'\int d^{2}y(<\sigma(0)\varepsilon(y)\varepsilon(y')
\sigma(\infty)>)^{2}\nn\\
&&\quad =\int_{1<|y'|<a}2\pi|y'|d|y'|\int_{1<|y|<|y'|}2\pi|y|d|y|
\frac{D^{4}}{|y|^{2\Delta_{\varepsilon}}
|y'|^{2\Delta_{\varepsilon}}}+\cdots\nn\\
&&\quad =4\pi^{2}D^{4}\int_{1<|y'|<a} d|y'||y'|^{-1+3\epsilon}
\int_{1<|y|<|y'|}d|y||y|
^{-1+3\epsilon}+\cdots\nn\\
&&\quad =4\pi^{2}D^{2}\int_{1<|y'|<a}d|y'|^{-1+3\epsilon}\frac{1}{3\epsilon}|
y'|^{3\epsilon}+\cdots\nn\\
&&\quad =2\pi^{2}D^{2}\frac{1}{9\epsilon^{2}}a^{6\epsilon}+\cdots
\eea
We have used: $<\sigma(0)\sigma(\infty)>=1,\,\,2\Delta_{\varepsilon}=
2-3\epsilon,$ and we have kept the 
leading term when integrating: 
$|y'|^{3\epsilon}-1\approx |y'|^{3\epsilon},\,\,\,a^{6\epsilon}-1
\approx a^{6\epsilon}.$

Adding to (A.34) the $1/\epsilon^{2}$ piece in (A.36), multiplied by
two because of two equivalent configurations, (A.35) and (A.36), one
finally obtains:
\beq
I_{3}=(4\pi^{2}D^{4}\frac{1}{9\epsilon^{2}}-\frac{\pi^{2}}{6\epsilon})
a^{6\epsilon}+O(1)
\eeq

The leading singularity, $\sim1/\epsilon^{2}$, is missed in the
calculation of the integral $I$ in the Appendix B for the following
reason. When $I_{3}$ is put in the form of (A.16), (A.17), the factor
$1/\epsilon$ is already in front, $1/\epsilon^{2}$ singularity of
$I_{3}$ would be produced by $1/\epsilon$ singularity of
$\tilde{I}_{3}$, or $I$, eq.(A.31).  Instead we find the result (A.32)
for $I,\,\,I\sim 1.$ In the integral $\tilde{I}_{3}$,(A.17), or in
$I,$ (A.31), the integration over $y$ is performed over the whole
infinite plane.  This is correct to define the finite piece of the
integral, but the singularity $1/\epsilon$ get cancelled in this way.

In fact, let us consider the integral $\tilde{I}_{3}$, (A.17). There
are two configurations which lead to $1/\epsilon$ singularity in the
integral (A.17):
\beq
0<|y|\ll1
\eeq
and
\beq
1\ll|y|<\infty
\eeq
For the first one we use:
\beq
\sigma(0)\varepsilon(y)=\frac{D}{|y|^{\Delta_{\varepsilon}}}\sigma(0)+\cdots
\eeq
and we get:
\beq
<\sigma(0)\varepsilon(y)\varepsilon(1)\sigma(\infty)> =
\frac{D}{|y|^{\Delta_{\varepsilon}}}<\sigma(0)\varepsilon(1)\sigma(\infty)>=
\frac{D^{2}}{|y|^{\Delta_{\varepsilon}}}
\eeq
We have used $<\sigma(0)\varepsilon(1)\sigma(\infty)>=D.$ For the 
second configuration, (A.41), we use:
\beq
\sigma(0)\varepsilon(1)=D\sigma(0)+\cdots
\eeq
and we get:
\bea
<\sigma(0)\varepsilon(y)\varepsilon(1)\sigma(\infty)>
&=&<\sigma(0)\varepsilon(1)\varepsilon(y)\sigma(\infty)>\nn\\
&=&D<\sigma(0)\varepsilon(y)\sigma(\infty)>\nn\\
&=&D\frac{1}{|y|^{\Delta_{\varepsilon}}}<\sigma(0)\varepsilon(1)\sigma(\infty)>
=\frac{D^{2}} {|y|^{\Delta_{\varepsilon}}}
\eea
In the last line we have used the scaling properties of the function 
$<\sigma(0)\varepsilon(y)\sigma(\infty)>.$ In the integral $\tilde{I}_{3}$, 
(A.17), the above two configurations will provide the following contribution:
\bea
\tilde{I}_{3}&=&\int_{0<|y|\ll1}d^{2}y\frac{D^{4}}{|y|^{2\Delta_{\varepsilon}}}+
\int_{1\ll|y|<\infty}d^{2}y\frac{D^{4}}{|y|^{2\Delta_{\varepsilon}}}+\cdots\nn\\
&=& 2\pi\int_{0<|y|\ll1}d|y||y|^{-1+3\epsilon}+2\pi\int_{1\ll|y|
<\infty}d|y||y|^{-1+3\epsilon}+\cdots
\eea
Next we change the variable in the second integral: $|y|\rightarrow
1/|y|$. Then we obtain:
\beq
\tilde{I}_{3}=2\pi\int_{0<|y|\ll1}d|y||y|^{-1+3\epsilon}+2\pi\int_{0<|y|\ll1} 
d|y||y|^{-1-3\epsilon}+\cdots
\eeq
If the first integral is $\sim 2\pi/3\epsilon$ (by extending the
integration up to 1), then, by analytic continuation in $\epsilon$,
the second integral should be defined to be $\sim-2\pi/3\epsilon$.  In
the result the two contributions $\sim 1/\epsilon$ cancel one
another. This is what is happening with the $1/\varepsilon$ terms in
the calculation of the integral $I$ in the Appendix B. The calculation
there is all based on the analytic continuation technique.

\section{Appendix}

The integral $I$, eq.(A.31). The integral is of the following general form:
\bea
I&=& \int d^{2}t\int d^{2}x\int d^{2}y|t|^{2\alpha'}|t-1|^{2\beta'}\nn\\
&& \quad \times
|x|^{2\alpha}|x-1|^{2\beta}|x-t|^{2\rho}|y|^{2\alpha}|y-1|^{2\beta}|y-t|^{2\rho}
\eea
We shall calculate it for the values of exponents 
$\alpha',\beta',\alpha,\beta,\rho $
which will be specified below. By the standard technique, see e.g.[12], 
this integral could be factorized into the following sum of products 
of contour integrals:
\bea
I&=& -\{
j^{(+)}_{1}[s(\beta')s^{2}(\beta)j^{(-)}_{1}+s(\beta')s(\beta)s(\beta+\rho)
j^{(-)}_{2}+s(\beta')s^{2}(\beta+\rho)j^{(-)}_{3}]\nn\\
&&\quad + 
j^{(+)}_{2}[s(\beta'+\rho)s^{2} (\beta)j^{(-)}_{1}
+\frac{1}{2}(s(\beta')s^{2}(\beta)+s(\beta'+\rho)s(\beta)s(\beta+\rho))
j^{(-)}_{2}\nn\\
&&\qquad \qquad +
s(\beta')s(\beta)s(\beta+\rho)j^{(-)}_{3}]\nn\\
&&\quad +
j^{(+)}_{3}[s(\beta'+2\rho) s^{2}(\beta)j^{(-)}_{1}
+s(\beta'+\rho)s^{2}(\beta)j^{(-)}_{2}+s(\beta')s^{2}
(\beta)j^{(-)}_{3}]\}
\eea
Here $s(\beta)\equiv\sin\pi\beta$, etc. The contour integrals are defined 
in the following way:
\bea
j^{(+)}_{1}&=&\int^{1}_{0}dt\int^{t}_{0}dx
\int^{t}_{0}dy(t)^{\alpha'}(1-t)^{\beta'}
\nn\\&&\qquad \times 
(x)^{\alpha}(1-x)^{\beta}(t-x)^{\rho}(y)^{\alpha}(1-y)^{\beta}(t-y)^{\rho}\\
j^{(+)}_{2}&=&2\int^{1}_{0}dt\int^{1}_{t}dx\int^{t}_{0}dy(\cdots)\\
j^{(+)}_{3}&=&\int^{1}_{0}dt\int^{1}_{t}dx\int^{1}_{t}dy(\cdots)
\eea
The symbol ($\cdots$) in (B.4), (B.5) stands for the same expression as in (B.3) 
except that the variables are put in the order corresponding to the order 
of integration, so that the differences of variables are always positive. 
E.g. the factors $(t-x)^{\rho},(t-y)^{\rho}$ in (B.3) will be in the 
form $(x-t)^{\rho},(x-t)^{\rho}$ in the integral (B.5).
\bea
j^{(-)}_{1}&=&\int^{\infty}_{1}dt\int^{t}_{1}dx\int^{t}_{1}dy(t)^{\alpha'}
(t-1)^{\beta'}\nn\\
&& \qquad \times(x)^{\alpha}(x-1)^{\beta}(t-x)^{\rho}(y)^{\alpha}
(y-1)^{\beta}(t-y)^{\rho}\\
j^{(-)}_{2}&=& 2\int^{\infty}_{1}dt\int^{\infty}_{t}dx\int^{t}_{1}dy(\cdots)\\
j^{(-)}_{3}&=& \int_{1}^{\infty}dt\int^{\infty}_{t}dx\int^{\infty}_{t}dy(\cdots)
\eea
The integrals $j^{(-)}_{1},j^{(-)}_{2},j^{(-)}_{3}$ can be put in the 
same form as the
integrals $j^{(+)}_{1}, j^{(+)}_{2}, j^{(+)}_{3}$ under the change of 
variables $t\rightarrow 1/t,\,\,x\rightarrow 1/x,\,\,y\rightarrow 1/y$. 
One obtains:
\bea
j^{(-)}_{1}&=&\int^{1}_{0}dt\int^{1}_{t}dx\int^{1}_{t}dy(t)^{\tilde{\alpha}'}
(1-t)^{\beta'}\nn\\
&&\qquad \times(x)^{\tilde{\alpha}}(1-x)^{\beta}(x-t)^{\rho}
(y)^{\tilde{\alpha}} (1-y)^{\beta} (y-t)^{\rho} \\
j^{(-)}_{2}&=&2\int^{1}_{0}dt\int^{t}_{0}dx\int^{1}_{t}dy(\cdots)\\
j^{(-)}_{3}&=&\int^{1}_{0}dt\int^{t}_{0}dx\int^{t}_{0}dy(\cdots)
\eea
Here
\beq
\tilde{\alpha}'=-2-\alpha'-\beta'-2\rho,\,\,\,\tilde{\alpha}
=-2-\alpha-\beta-\rho
\eeq
Originally the values of the exponents are given by:
\bea
\alpha' &=& 4\bar{\alpha}_{\sigma}\alpha_{\varepsilon}
=\frac{1}{3}-\frac{\epsilon}{2}\\
\beta' &=&4\alpha^{2}_{\varepsilon}=\frac{4}{3}+\epsilon\\
\alpha &=&2\bar{\alpha}_{\sigma}\alpha_{+}=-\frac{1}{3}+\frac{\epsilon}{2}\\
\beta &=&2\alpha_{\varepsilon}\alpha_{+}=-\frac{4}{3}-\epsilon\\
\rho &=&2\alpha_{\varepsilon}\alpha_{+}=-\frac{4}{3}-\epsilon
\eea
Here $\bar{\alpha}_{\sigma},\,\alpha_{\varepsilon}$ correspond to the 
Coulomb gas operators
$$
V_{\bar{\alpha}_{\sigma}}(x)=:\mbox{exp}\{i\bar{\alpha}_{\sigma}\varphi(x)\}: 
\quad ; \quad
V_{\alpha_{\varepsilon}}(x)=:\mbox{exp}\{i\alpha_{\varepsilon}\varphi(x)\}:
$$
which represent the operators of spin $\sigma(x)$ and energy 
$\varepsilon(x);\,\,\alpha_{+}$ corresponds to the screening
operator. In particular:
\beq
\alpha_{\varepsilon}=\alpha_{1.2}=-\frac{\alpha_{+}}{2}
\eeq
For the spin operator, if $\alpha^{2}_{+}$ is put in the form:
\beq
\alpha^{2}_{+}=\frac{2p}{2p-1}=\frac{4}{3}+\epsilon
\eeq
(which corresponds to the Kac table of the size $(2p-1)\times(2p-2)$) 
then the Potts model spin operator is
\beq
\sigma\sim V_{p,p-1},V_{\overline{p,p-1}}
\eeq
$V_{\overline{p,p-1}}(x)$ is the conjugate Coulomb gas operator 
(with respect to $V_{p,p-1}(x)$), which we are actually using:
\beq
V_{\overline{p,p-1}}(x)=:\mbox{exp}\{i\alpha_{\overline{p,p-1}}\varphi(x)\}:
\eeq
\beq
\alpha_{\overline{p,p-1}}\equiv\bar{\alpha}_{\sigma}=
\frac{1+p}{2}\alpha_{-}+\frac{p}{2}\alpha_{+}
\eeq
One gets the values of exponents in (B.13)-(B.17), using in addition 
the relation between $p$ and
$\alpha^{2}_{+}$:
\beq
p(\alpha^{2}_{+}-1)=\frac{\alpha^{2}_{+}}{2}
\eeq
which follows from (B.19), and the usual relation $\alpha_{+}\alpha_{-}=-1$ 
for the Coulomb gas parameters.

We have found that the calculation of the integral $I$ is simpler 
for the values of exponents $\alpha', \beta', \alpha, \beta, \rho$ 
obtained after the transformation of variables:
\beq
t\rightarrow 1-\frac{1}{t},\,\,x\rightarrow 1-\frac{1}{x},\,\,y
\rightarrow 1-\frac{1}{y}
\eeq
One gets then the following values:
\beq
\alpha'=\frac{4}{3}+\epsilon,\,\,\,\beta'=-1+\frac{3}{2}\epsilon
\eeq
\beq
\alpha=-\frac{4}{3}-\epsilon,\,\,\,\beta=1+\frac{3}{2}\epsilon
\eeq
\beq
\rho=-\frac{4}{3}-\epsilon
\eeq
The exponents $\tilde{\alpha}', \tilde{\alpha}$ of the contour 
integrals $j^{(-)}_{1}, j^{(-)}_{2}, j^{(-)}_{3}$ eqs. (B.9)-(B.11), 
are obtained from (B.12):
\beq
\bar{\alpha}'=\frac{1}{3}-\frac{\epsilon}{2}
\eeq
\beq
\tilde{\alpha}=-\frac{1}{3}+\frac{\epsilon}{2}
\eeq

We shall give next some details on the calculation of the contour integrals 
$j^{(+)}_{1}, j^{(+)}_{2}, j^{(+)}_{3},$ which enter into the decomposition 
of the integral $I$, eq.(B.2). The calculation of the integrals 
$j^{(-)}_{3}, j^{(-)}_{2}, j^{(-)}_{1}$ is respectively analogous.

\underline{$j^{(+)}_{1}, j^{(-)}_{3}.$}
\bea
j^{(+)}_{1}&=&\int^{1}_{0}dt\,\,t^{\alpha'}(1-t)^{\beta'}
\int^{t}_{0}dx\,\,x^{\alpha}(1-x)^{\beta}(t-x)^{\rho}\nn\\
&&\qquad \times\int^{t}_{0}dy\,\,y^{\alpha}(1-y)^{\beta}(t-y)^{\rho}
\eea
We change variables: $x=\tilde{x}t,\,\,y=\tilde{y}t$ and next we 
drop the tildes of the new variables $\tilde{x},\,\tilde{y}.$
\bea
j^{(+)}_{1}&=&\int^{1}_{0}dt\,\,t^{2+\alpha'+2\alpha+2\rho}(1-t)^{\beta'}
\int^{1}_{0}dx\,\,x^{\alpha}(1-x)^{\rho}(1-xt)^{\beta}\nn\\
&&\qquad \times \int^{1}_{0}dy\,\,y^{\alpha}(1-y)^{\rho}(1-yt)^{\beta}
\eea
We expand next the factors $(1-xt)^{\beta},\,(1-yt)^{\beta}:$
\beq
(1-xt)^{\beta}=\sum^{\infty}_{k=0}\frac{(-\beta)_{k}}{k!}(xt)^{k}
\eeq
Here $(-\beta)_{k}=(-\beta)(-\beta+1)\cdots(-\beta+k-1)$. One obtains:
\bea
j^{(+)}_{1}&=&\sum_{k_{1}}\sum_{k_{2}}
\frac{(-\beta)_{k_{1}}}{k_{1}!}\frac{(-\beta)_{k_{2}}}{k_{2}!}
\int^{1}_{0}dt\,\,t^{2+\alpha'+2\alpha+2\rho+k_{1}
+k_{2}}(1-t)^{\beta'}\nn\\
&&\qquad \times\int^{1}_{0}dx\,\, x^{\alpha+k_{1}}(1-x)^{\rho}
\int^{1}_{0}dy\,\,y^{\alpha+k_{2}}(1-y)^{\rho}
\eea
Next we use the integral
\beq
\int^{1}_{0}dt\,\,t^{a}(1-t)^{b}=\frac{\Gamma(1+a)\Gamma(1+b)}{\Gamma(2+a+b)}
\eeq
to obtain:
\bea
j^{(+)}_{1}&=&\sum_{k_{1}}\sum_{k_{2}}\frac{(-\beta)_{k_{1}}}{k_{1}!}
\frac{(-\beta)_{k_{2}}}{k_{2}!}\frac{\Gamma(3+\alpha'
+2\alpha+2\rho+k_{1}+k_{2})\Gamma(1+\beta')}{\Gamma(4+\alpha'
+\beta'+2\alpha+2\rho+k_{1}+k_{2})}\nn\\
&&\qquad \times\frac{\Gamma(1+\alpha+k_{1})\Gamma(1+\rho)}
{\Gamma(2+\alpha+\rho+k_{1})}\frac{\Gamma(1+\alpha
+k_{2})\Gamma(1+\rho)}{\Gamma(2+\alpha+\rho+k_{2})} 
\eea
By using repeatedly the recurrence relation for the 
$\Gamma$-function, $\Gamma(z+1)=z\Gamma(z)$,
one can put (B.35) into the following form:
\bea
j^{(+)}_{1}&=&\gamma^{(+)}_{1}\times S^{(+)}_{1} \\
\gamma^{(+)}_{1}&=&\frac{\Gamma(3+\alpha'+2\alpha+2\rho)\Gamma(1
+\beta')}{\Gamma(4+\alpha'+\beta'+2\alpha+2\rho)}(\frac{\Gamma(1
+\alpha)\Gamma(1+\rho)}{\Gamma(2+\alpha+\rho)})^{2}\\
S^{(+)}_{1}&=&\sum_{k_{1}}\sum_{2}\frac{(-\beta)_{k_{1}}}{k_{1}!}
\frac{(-\beta)_{k_{2}}}{k_{2}!}\frac{(3+\alpha'+2\alpha+2\rho)_{k_{1}
+k_{2}}}{(4+\alpha'+\beta'+2\alpha+2\rho)_{k_{1}+k_{2}}}\nn\\
&& \qquad \times\frac{(1+\alpha)_{k_{1}}}{(2+\alpha+\rho)_{k_{1}}}
\frac{(1+\alpha)_{k_{2}}}{(2+\alpha+\rho)_{k_{2}}}
\eea
Substituting the values of exponents (B.25)-(B.27) one obtains:
\beq
\gamma^{(+)}_{1}=\frac{\Gamma(-1-3\epsilon)
\Gamma(\frac{3}{2}\epsilon)}{\Gamma(-1-\frac{3}{2}\epsilon)}
(\frac{\Gamma(-\frac{1}{3}-\epsilon)\Gamma(-\frac{1}{3}
-\epsilon)}{\Gamma(-\frac{2}{3}-2\epsilon)})^{2}
\eeq
\beq
S^{(+)}_{1}=\sum_{k_{1}}\sum_{k_{2}}\frac{(-1
-\frac{3}{2}\epsilon)_{k_{1}}}{k_{1}!}\frac{(-1-\frac{3}{2}
\epsilon)_{k_{2}}}{k_{2}!}\frac{(-\frac{1}{3}-\epsilon)_{k_{1}}}{(-\frac{2}{3}
-2\epsilon)_{k_{1}}}
\frac{(-\frac{1}{3}-\epsilon)_{k_{2}}}{(-\frac{2}{3}
-2\epsilon)_{k_{2}}}\frac{(-1-3\epsilon)_{k_{1}+
k_{2}}}{(-1-\frac{3}{2}\epsilon)_{k_{1}+k_{2}}}
\eeq
We shall do calculations by expanding in $\epsilon$. For the
renormalization group equation we need to keep the first two terms
only.

For $\gamma^{(+)}_{1}$ in (B.39) a simple calculation gives:
\beq
\gamma^{(+)}_{1}=\frac{\gamma^{2}}{3\epsilon}(1-\frac{3}{2}\epsilon
-4\epsilon\kappa)+O(\epsilon)
\eeq
We have defined here:
\beq
\gamma=\frac{(\Gamma(-\frac{1}{3}))^{2}}{\Gamma(-\frac{2}{3})}
\eeq
\beq
\kappa=\psi(-\frac{1}{3})-\psi(-\frac{2}{3})=\frac{\pi}{\sqrt{3}}+\frac{3}{2}
\eeq
$\psi(z)$ is the standard $\psi$-function:
\beq
\psi(z)=\frac{d}{dz}\log\Gamma(z)
\eeq

The calculation of $S^{(+)}_{1}$ is more involved. First we check the
convergence of the series in (B.40). In general one has:
\beq
\frac{(a)_{k}}{(b)_{k}}\approx (k)^{a-b},\,\,\,k\gg 1
\eeq
Using this asymptotic form one gets the following estimate for large 
$k_{1}, k_{2}$ in (B.40):
\beq
S^{(+)}_{1}\sim\sum_{k_{1}}\sum_{k_{2}}(k_{1})^{-\frac{5}{3}
-\frac{\epsilon}{2}}(k_{2})
^{-\frac{5}{3}-\frac{\epsilon}{2}}(k_{1}+k_{2})^{-\frac{3}{2}\epsilon}
\eeq
(We observe that $k!=(1)_{k})$. So the series converge and we can do 
safely its $\epsilon$-expansion.

Using the specific values of the parameters in the series (B.40) one
can develop it in the following way:
\beq
S^{(t)}_{1}\approx(0,0)+2(0,1)+(1,1)+2\sum^{\infty}_{k_{2}=2(k_{1}=0)}(\cdots)
+2\sum^{\infty}
_{k_{2}=2(k_{1}=1)}(\cdots)
\eeq
(0,0) stand for the term of the series $k_{1}=0, k_{2}=0$, etc. We 
have dropped the part of the series corresponding to 
$(k_{1}=2,\cdots\infty, k_{2}=2,\cdots,\infty)$ since it is 
$\sim\epsilon^{2}$. We are keeping only terms $\sim 1$ and 
$\sim\epsilon$ in $S^{(+)}_{1}$. Explicitly one gets:
\bea
S^{(+)}_{1}&\approx& 1+2\frac{(-1-\frac{3}{2}\epsilon)}{1}\frac{(-\frac{1}{3}
-\epsilon)}{(-\frac{2}{3}-2\epsilon)}\frac{(-1-3\epsilon)}{(-1
-\frac{3}{2}\epsilon)}\\
&& +\frac{(-1-\frac{3}{2}\epsilon)^{2}}{1}(\frac{-\frac{1}{3}
-\epsilon}{-\frac{2}{3}-2\epsilon})^{2}\frac{(-1
-3\epsilon)(-3\epsilon)}{(-1-\frac{3}{2}\epsilon)(-\frac{3}{2}\epsilon)}\nn\\
&&+2\sum_{k=0}^{\infty}\frac{(-1-\frac{3}{2}\epsilon)_{k+2}}{(k
+2)!}\frac{(-\frac{1}{3}-\epsilon)_{k+2}}{(-\frac{2}{3}
-2\epsilon)_{k+2}}\frac{(-1-3\epsilon)_{k+2}}{(-1-\frac{3}{2}\epsilon)_{k
+2}}\nn\\
&&+2\sum_{k=0}^{\infty}\frac{(-1-\frac{3}{2}\epsilon)}{1}\frac{(-1
-\frac{3}{2}\epsilon)_{k+2}}{(k+2)!}\frac{(-\frac{1}{3}-\epsilon)}{(-\frac{2}{3}
-2\epsilon)}\frac{(-\frac{1}{3}-\epsilon)_{k+2}}{(-\frac{2}{3}-2\epsilon)_{k
+2}}\frac{(-1-3\epsilon)_{k+3}}{(-1-\frac{3}{2}\epsilon)_{k+3}}\nn
\eea
To simplify further, we use the following relations:
\bea
&&(a)_{k+2}=a(a+1)(a+2)_{k},\nn\\
&& (a)_{k+3}=a(a+1)(a+2)(a+3)_{k}
\eea
In particular:
\beq
(-1-\frac{3}{2}\epsilon)_{k+2}=(-1-\frac{3}{2}\epsilon)(-\frac{3}{2}\epsilon)
(1-\frac{3}{2}\epsilon)_{k}\approx\frac{3}{2}\epsilon(1)_{k}
=\frac{3}{2}\epsilon k!
\eeq
\beq
(-\frac{1}{3}-\epsilon)_{k+2}\approx(-\frac{1}{3})_{k+2}=(-\frac{1}{3})(\frac{2}{3})(\frac{5}{3})_{k}
\eeq
etc.  Using these simple rules, after some algebra one gets $S^{(+)}_{1}$ in the following form:
\beq
S^{(+)}_{1}=\frac{1}{2}+\epsilon(3s-\frac{3}{4})+O(\epsilon^{2})
\eeq
where
\beq
s=\sum^{\infty}_{k=0}\frac{k!}{(k+2)!}\frac{(\frac{5}{3})_{k}}{(\frac{4}{3})_{k}}
\eeq
It remains to calculate this sum. We shall use the following general results:
\beq
\sum^{\infty}_{k=0}\frac{(a)_{k}(b)_{k}}{k!(c)_{k}}=\frac{\Gamma(c)\Gamma(c-a-b)}{\Gamma
(c-a)\Gamma(c-b)}
\eeq
\beq
\sum^{\infty}_{k=0}\frac{k!}{(k+2)!}\frac{(b)_{k}}{(c)_{k}}=\frac{c-1}{b-1}(\frac{c-1}{c-2}+
\frac{c-b}{b-2}\tilde{\kappa})
\eeq
Here
\beq
\tilde{\kappa}=\psi(c-b)-\psi(c-2)
\eeq
Eq. (B.55) could be derived by using the sum (B.54), which is standard, 
and some rather simple algebra.

From (B.55) and (B.53) one gets in a straightforward way:
\beq
s=-\frac{1}{4}+\frac{\kappa}{2}
\eeq
The constant $\kappa$ is defined in (B.43). By eq.(B.52)
\beq
S^{(+)}_{1}\approx\frac{1}{2}+\epsilon(3s-\frac{3}{4})
=\frac{1}{2}-\frac{3}{2}\epsilon+\frac{3}{2}\epsilon\kappa
\eeq
Returning still back to eqs. (B.36), (B.41) one obtains:
\beq
j^{(+)}_{1}=\gamma^{(+)}_{1}S^{(+)}_{1}=\frac{\gamma^{2}}{6\epsilon}(1-\frac{9}{2}\epsilon
-\kappa\epsilon)+O(\epsilon)
\eeq
The integral $j^{(-)}_{3}$, eq.(B.11), is of the same form as $j^{(+)}_{1}$ with only the
exponents $\alpha',\alpha$ replaced by $\tilde{\alpha'},\tilde{\alpha}$, eqs.(B.28), 
(B.29).
The calculation follows the same lines, with the result:
\beq
j^{(-)}_{3}=\gamma^{2}(-\frac{7}{4}+\frac{\kappa}{2})+O(\epsilon)
\eeq

\underline{$j^{(+)}_{2}, j^{(-)}_{2}.$}

For calculation of $j^{(+)}_{2}$ it is useful to use the following linear relation 
of the integrals
\beq
u_{2}=-\frac{s(\alpha)}{s(\alpha+\rho)}u_{1}-\frac{s(\alpha+\beta+\rho)}{s(\alpha
+\rho)}\tilde{u}_{2}
\eeq
Here we have redefined, for the purpose of this particular calculation only,
\beq
u_{2}=\frac{1}{2}j^{(+)}_{2},\,\,\,u_{1}=j^{(+)}_{1}
\eeq
and $\tilde{u}_{2}$ is a  new integral:
\bea
\tilde{u}_{2}=\int^{1}_{0}dt\,\,t^{\alpha'}(1-t)^{\beta'}\int^{t}_{0}dy\,\,y^{\alpha}
(1-y)^{\beta}(t-y)
\int^{\infty}_{1}dx\,\,x^{\alpha}(x-1)^{\beta}(x-t)^{\rho}
\eea
$s(\alpha)$ in (B.61) is $\sin\pi\alpha.$ The relation (B.61) is obtained in a standard 
way by doing transformation of the contours of integration [12].

Putting the problem this way, to calculate $j^{(+)}_{2}$ we have to
calculate the integral $\tilde{u}_{2}$, (B.63). Because of specific
values of the exponents the calculation of the integral
$\tilde{u}_{2}$ is simpler, as compared to the direct calculation of
$j^{(+)}_{2}$.

We do first the change of variables in (B.63): $x\rightarrow 1/x,\,\,y\rightarrow ty$.
One gets:
\bea
\tilde{u}_{2}&=&\int^{1}_{0}dt\,\,t^{1+\alpha'+\alpha+\rho}(1-t)^{\beta'}
\int^{1}_{0}dy\,\,y^{\alpha}(1-y)^{\rho}(1-ty)^{\beta}\nn \\
&& \qquad \times
\int^{1}_{0}dx\,\,x^{\tilde{\alpha}} (1-x)^{\beta}(1-xt)^{\rho}
\eea
$\tilde{\alpha}$ was defined in (B.12), (B.29). Next we expand the factors 
$(1-ty)^{\beta}, (1-xt)^{\rho}$ and proceed like we did for the integral 
$j^{(+)}_{1}$, starting with eq.(B.31). We get $\tilde{u}_{2}$ in the 
following form:
\beq
\tilde{u}_{2}=\tilde{\gamma}_{2}\tilde{S}_{2}
\eeq
\beq
\tilde{\gamma}_{2}=\frac{\Gamma(2+\alpha'+\alpha+\rho)\Gamma(1+\beta')}{\Gamma(3+\alpha'+\beta'+
\alpha+\rho)}\frac{\Gamma(1+\alpha)\Gamma(1+\rho)}{\Gamma(2+\alpha+\rho)}\frac{\Gamma(1+
\tilde{\alpha})\Gamma(1+\beta)}{\Gamma(2+\tilde{\alpha}+\beta)}
\eeq
\beq
\tilde{S}_{2}=\sum_{k}\sum_{l}\frac{(-\beta)_{k}}{k!}\frac{(-\rho)_{l}}{l!}\frac{(1+\alpha)_{k}}
{(2+\alpha+\rho)_{k}}\frac{(1+\tilde{\alpha})_{l}}{(2+\tilde{\alpha}+\beta)_{l}}\frac{(2+\alpha'
+\alpha+\rho)_{k+l}}{(3+\alpha'+\beta'+\alpha+\rho)_{k+l}}
\eeq
Substituting the values of the parameters, eqs.(B.25)-(B.29), one obtains:
\beq
\tilde{\gamma}_{2}=\frac{\Gamma(\frac{2}{3}-\epsilon)\Gamma(\frac{3}{2}\epsilon)}{\Gamma(\frac{2}{3}+
\frac{\epsilon}{2})}\frac{\Gamma^{2}(-\frac{1}{3}-\epsilon)}{\Gamma(-\frac{2}{3}-2\epsilon)}\frac
{\Gamma(\frac{2}{3}+\frac{\epsilon}{2})\Gamma(2+\frac{3}{2}\epsilon)}{\Gamma(\frac{8}{3}+2\epsilon)}
\eeq
\beq
\tilde{S}_{2}=\sum_{k}\sum_{l}\frac{(-1-\frac{3}{2}\epsilon)_{k}}{k!}\frac{(\frac{4}{3}+\epsilon)_{l}}
{l!}\frac{(-\frac{1}{3}-\epsilon)_{k}}{(-\frac{2}{3}-2\epsilon)_{k}}\frac{(\frac{2}{3}+\frac{\epsilon}
{2})_{l}}{(\frac{8}{3}+2\epsilon)_{l}}\frac{(\frac{2}{3}-\epsilon)_{k+l}}{(\frac{2}{3}+\frac{\epsilon}
{2})_{k+l}}
\eeq
For $\tilde{\gamma}_{2}$ one gets:
\beq
\tilde{\gamma}_{2}=\frac{3}{5\epsilon}\gamma(1+(\frac{15}{2}-\frac{6}{5})\epsilon-2\epsilon\kappa
-3\epsilon\psi(-\frac{1}{3})+3\epsilon\psi(1))+O(\epsilon)
\eeq
For $\tilde{S}_{2}$ one gets asymptotically, for large $k,l$:
\beq
\tilde{S}_{2}\sim\sum_{k}\sum_{l}(k)^{-\frac{5}{3}-\frac{\epsilon}{2}}(l)^{-\frac{5}{3}-
\frac{\epsilon}{2}}(k+l)^{-\frac{3}{2}\epsilon}
\eeq
-- the series is convergent. In order to develop in $\epsilon, \tilde{S}_{2}$ 
can be decomposed
in the following way:
\beq
\tilde{S}_{2}=(0,l)+(1,l)+(k+2,l)
\eeq
or explicitly:
\bea
\tilde{S}_{2}&\approx& \sum^{\infty}_{l=0}\frac{(\frac{4}{3}
+\epsilon)_{l}}{l!}\frac{(\frac{2}{3}+\frac{\epsilon}{2})_{l}}{(\frac{8}{3}
+2\epsilon)_{l}}\frac{(\frac{2}{3}-\epsilon)_{l}}{(\frac{2}{3}
+\frac{\epsilon}{2})_{l}}\nn\\
&&+\frac{(-1-\frac{3}{2}\epsilon)}{1}\frac{(-\frac{1}{3}-\epsilon)}
{(-\frac{2}{3}-2\epsilon)}\frac{(\frac{2}{3}-\epsilon)}{(\frac{2}{3}
+\frac{\epsilon}{2})}\sum^{\infty}_{l=0}\frac{(\frac{4}{3}
+\epsilon)_{l}}{l!}\frac{(\frac{2}{3}+\frac{\epsilon}{2})_{l}}
{(\frac{8}{3}+2\epsilon)_{l}}\frac{(\frac{5}{3}-\epsilon)_{l}}
{(\frac{5}{3}+\frac{\epsilon}{2})_{l}}\nn\\
&&+(-1)(-\frac{3}{2}\epsilon)\frac{(-\frac{1}{3})(\frac{2}{3})}{(-\frac{2}{3})
(\frac{1}{3})}\sum^{\infty}_{k=0}\frac{k!}{(k+2)!}\frac{(\frac{5}{3})_{k}}{(
\frac{4}{3})_{k}}\sum^{\infty}_{l=0}\frac{(\frac{4}{3})_{l}}{l!}
\frac{(\frac{2}{3})_{l}}{(\frac{8}{3})_{l}}
\eea
In the last term we have put $\epsilon=0$ because of the factor 
$(-\frac{3}{2}\epsilon)$ in front. Finally one gets:
\beq
\tilde{S}_{2}\approx s_{1}-\frac{1}{2}(1-\frac{3}{4}\epsilon)s_{2}
+\frac{3}{2}\epsilon s_{3}s_{4}
\eeq
with 
\bea
s_{1}&=&\sum^{\infty}_{l=0}\frac{(\frac{4}{3}+\epsilon)_{l}(\frac{2}{3}
-\epsilon)_{l}}{l!(\frac{8}{3}+2\epsilon)_{l}}\\
s_{2}&=&\sum^{\infty}_{l=0}\frac{(\frac{4}{3}+\epsilon)_{l}}{l!}
\frac{(\frac{2}{3}+\frac{\epsilon}{2})_{l}}{(\frac{8}{3}
+2\epsilon)_{l}}\frac{(\frac{5}{3}-\epsilon)_{l}}{(\frac{5}{3}+\frac
{\epsilon}{2})_{l}} \\
s_{3}&=&\sum^{\infty}_{k=0}\frac{k!}{(k+2)!}\frac{(\frac{5}{3})_{k}}{(\frac{4}{3})_{k}}\\
s_{4}&=&\sum^{\infty}_{l=0}\frac{(\frac{4}{3})_{l}}{l!}\frac{(\frac{2}{3})_{l}}{(
\frac{8}{3})_{l}}
\eea
By using the results for sums in (B.54), (B.55) the calculation of 
$s_{1}, s_{3}, s_{4}$ is straightforward. One obtains:
\bea
s_{1}&=&-\frac{5}{9}(1+(\frac{6}{5}-\frac{27}{2})\epsilon
+\kappa\epsilon+3\epsilon\psi(-\frac{1}{3})-3\epsilon\psi(1))
+O(\epsilon^{2})\\
s_{3}&=&-\frac{1}{4}+\frac{\kappa}{2}\\
s_{4}&=&-\frac{5}{9}\gamma
\eea
For the sum $s_{2}$, we observe at first that
\beq
\frac{(\frac{2}{3}+\frac{\epsilon}{2})_{l}}{(\frac{5}{3}+\frac{\epsilon}{2})_{l}}=
\frac{\frac{2}{3}+\frac{\epsilon}{2}}{\frac{2}{3}+\frac{\epsilon}{2}+l}
\eeq
So one gets:
\bea
s_{2}&=&(\frac{2}{3}+\frac{\epsilon}{2})\sum^{\infty}_{l=0}\frac{(\frac{4}{3}
+\epsilon)_{l}}{l!}\frac{(\frac{5}{3}-\epsilon)_{l}}{(\frac{8}{3}
+2\epsilon)_{l}}\frac{1}{\frac{2}{3}+\frac{\epsilon}{2}+l}\\
&=&\frac{\frac{2}{3}+\frac{\epsilon}{2}}{(\frac{2}{3}-\epsilon)}
\sum^{\infty}_{l=0}\frac{(\frac{4}{3}+\epsilon)_{l}}{l!}\frac{(\frac{2}{3}
-\epsilon)_{l}}{(\frac{8}{3}+2\epsilon)_{l}}\frac{\frac{2}{3}-\epsilon
+l}{\frac{2}{3}+\frac{\epsilon}{2}+l}\nn\\
&=&\frac{\frac{2}{3}+\frac{\epsilon}{2}}{\frac{2}{3}-\epsilon}\sum^{\infty}_{l=0}
\frac{(\frac{4}{3}+\epsilon)_{l}}{l!}\frac{(\frac{2}{3}-\epsilon)_{l}}{(\frac{8}{3}
+2\epsilon)_{l}}(1-\frac{3}{2}\epsilon\frac{1}{\frac{2}{3}+\frac{\epsilon}{2}+l})\nn\\
&\approx& (1+\frac{3}{4}\epsilon+\frac{3}{2}\epsilon)(\sum_{l}\frac{(\frac{4}{3}
+\epsilon)_{l}}{l!}\frac{(\frac{2}{3}-\epsilon)_{l}}{(\frac{8}{3}
+2\epsilon)_{l}}-\frac{3}{2}\epsilon
\frac{2}{3}\frac{5}{3}\sum_{l}\frac{(\frac{4}{3})_{l}}{l!}\frac{1}{(\frac{5}{3}+l)
(\frac{2}{3}+l)^{2}})\nn
\eea
Calculation of the first sum is straightforward. One obtains:
\beq
\sum_{l}\frac{(\frac{4}{3}+\epsilon)_{l}}{l!}\frac{(\frac{2}{3}-\epsilon)_{l}}
{(\frac{8}{3}+2\epsilon)_{l}}\approx -\frac{5}{9}\gamma(1+\frac{6}{5}\epsilon-\frac{27}{2}
\epsilon+\epsilon\kappa+3\epsilon\psi(-\frac{1}{3})-3\epsilon\psi(1))
\eeq
The second sum could be calculated by decomposing
\beq
\frac{1}{(\frac{5}{3}+l)(\frac{2}{3}+l)^{2}}=\frac{1}{(\frac{2}{3}+l)^{2}}-\frac{1}{\frac{2}{3}
+l}+\frac{1}{\frac{5}{3}+l}
\eeq
and using the formulas:
\bea
\sum^{\infty}_{k=0}\frac{(a)_{k}}{k!}\frac{1}{b+k}&=&\frac{\Gamma(1-a)\Gamma(b)}
{\Gamma(1-a+b)}\\
\sum^{\infty}_{k=0}\frac{(a)_{k}}{k!}\frac{1}{(b+k)^{2}}&=&\frac{\Gamma(1-a)\Gamma(b)}
{\Gamma(1-a+b)}(\psi(1-a+b)-\psi(b))
\eea
One gets in this way:
\beq
\sum_{l}\frac{(\frac{4}{3})_{l}}{l!}\frac{1}{(\frac{5}{3}+l)(\frac{2}{3}+l)^{2}}=
\frac{\gamma}{2}(\frac{5}{2}-\kappa)
\eeq
From eqs.(B.83), (B.84), (B.88) one obtains:
\beq
s_{2}=-\frac{5}{9}\gamma(1+\frac{6}{5}\epsilon-\frac{15}{2}\epsilon-\frac{\kappa}{2}
\epsilon+3\epsilon\psi(-\frac{1}{3})-3\epsilon\psi_{1})
\eeq
Putting together the results for $s_{1}, s_{2}, s_{3}, s_{4}$ into eq.(B.74) for 
$\tilde{S}_{2}$ one gets:
\beq
\tilde{S}_{2}=-\frac{5}{18}\gamma(1+\frac{6}{5}\epsilon-\frac{39}{2}\epsilon
+4\kappa\epsilon +3\epsilon\psi(-\frac{1}{3})-3\epsilon\psi(1))+O(\epsilon^{2})
\eeq
Returning back to the eqs.(B.65), (B.68) one derives:
\beq
\tilde{u}_{2}=-\frac{\gamma^{2}}{6\epsilon}(1-7\epsilon+2\kappa\epsilon)+O(\epsilon)
\eeq
Going still back to the relation (B.61), putting the values of the parameters and
substituting the value of $u_{1}$ obtained earlier, one finally derives:
\beq
j^{(+)}_{2}\equiv 2u_{2}=\frac{\gamma^{2}}{6}(\frac{1}{2}-3\kappa)+O(\epsilon)
\eeq

The integral $j^{(-)}_{2}$, eq.(B.10), is of the same form as $j^{(+)}_{2}$ with only the
exponents $\alpha', \alpha $ replaced by $\tilde{\alpha'}, \tilde{\alpha}$, eqs.(B.28),
(B.29). The calculation follows the same line. There is one additional detail in 
the calculation.
In the decomposition of the sum $\tilde{S}_{2}$, corresponding to the integral 
$j^{(-)}_{2}$, one gets, in particular, the sums:
\beq
s_{1}=\sum_{l}\frac{(\frac{4}{3}+\epsilon)_{l}}{l!}\frac{(-\frac{1}{3}-\epsilon)_{l}
(\frac{2}{3}-\epsilon)_{l}}{(\frac{5}{3}+\frac{6}{2})_{l}(\frac{2}{3}+\frac{\epsilon}{2})_{l}}
\eeq
\beq
s_{2}=\sum_{l}\frac{(\frac{4}{3}+\epsilon)_{l}}{l!}\frac{(-\frac{1}{3}-\epsilon)_{l}}
{(\frac{5}{3}+\frac{\epsilon}{2})_{l}}\frac{(\frac{5}{3}-\epsilon)_{l}}{(\frac{5}{3}
+\frac{\epsilon}{2})_{l}}
\eeq
The sum $s_{1}$ is transformed as follows:
\beq
s_{1}=(-\frac{1}{3}-\epsilon)(\frac{2}{3}+\frac{\epsilon}{2})\sum_{l}\frac{(\frac{4}{3}
+\epsilon)_{l}}{l!}(\frac{(\frac{2}{3}-\epsilon)_{l}}{(\frac{2}{3}+\epsilon)_{l}})^{2}
\frac{1}{(-\frac{1}{3}-\epsilon+l)(\frac{2}{3}+\frac{\epsilon}{2}+l)}
\eeq
and next relation is used:
\beq
(\frac{(\frac{2}{3}-\epsilon)_{l}}{(\frac{2}{3}+\frac{\epsilon}{2})_{l}})^{2}=
\frac{(\frac{2}{3}-2\epsilon)_{l}}{(\frac{2}{3}+\epsilon)_{l}}+O(\epsilon^{2})
\eeq
So, up to linear order in $\epsilon$, one obtains $s_{1}$ in the form:
\beq
s_{1}\simeq(-\frac{1}{3}-\epsilon)(\frac{2}{3}+\frac{\epsilon}{2})\sum_{l}\frac{(
\frac{4}{3}+\epsilon)_{l}}{l!}\frac{(\frac{2}{3}-2\epsilon)_{l}}{(\frac{2}{3}
+\epsilon)_{l}}\frac{1}{(-\frac{1}{3}-\epsilon+l)(\frac{2}{3}+\frac{\epsilon}{2}+l)}
\eeq
In the rest of the calculation one proceeds in the way analogous to the calculation of the sum

$s_{2}$ for the integral $j^{(+)}_{2}$, eq.(B.83). The sum $s_{2}$, of the integral $j^{(-)}_{2}$,
eq.(B.94), is dealt with in a similar manner. Finally one obtains the following result for
$j^{(-)}_{2}$:
\beq
j^{(-)}_{2}=-\frac{\gamma^{2}}{3\epsilon}(1-\frac{3}{2}\epsilon-\kappa\epsilon)+O(\epsilon)
\eeq

\underline{$j^{(+)}_{3}, j^{(-)}_{1}.$}

We redefine: $j^{(+)}_{3}\equiv u_{3}$, and we shall use the following linear relation
between the integrals:
\beq
u_{3}=A^{2}u_{1}+2AB\tilde{u}_{2}+B^{2}u_{1}^{\ast}
\eeq
Here $u_{1}=j^{(+)}_{1}, \tilde{u}_{2}$ is defined in (B.63), $u^{\ast}_{1}$ is 
a new integral:
\beq
u^{\ast}_{1}=\int^{1}_{0}dt\,\,t^{\alpha}(1-t)^{\beta}\int^{\infty}_{1}dx\,\,
x^{\alpha} (x-1)^{\beta}(x-t)^{\rho}
\int^{\infty}_{1}dy\,\,y^{\alpha}(y-1)^{\beta}
(y-t)^{\rho}
\eeq
$A$ and $B$ in (B.99) are coefficients:
\beq
A=\frac{-s(\alpha)}{s(\alpha+\rho)},\,\,\,B=\frac{-s(\alpha+\beta+\rho)}{s(\alpha+\rho)}
\eeq
The relation (B.99) is obtained by transformations of the contours of integration [12],
similar to the relation (B.61)

The integrals $u_{1}, \tilde{u}_{2}$ had already been calculated. 
To define $u_{3}(j^{(+)}_{3})$ we have to define $u^{\ast}_{1}$, 
which turns out to be easier.

We change the variables: $x\rightarrow 1/x,\,y\rightarrow1/y$. This gives:
\bea
u^{\ast}_{1}=\int^{1}_{0}dt\,\,t^{\alpha'}(1-t)^{\beta'}
\int^{1}_{0}dx\,\,x^{\tilde{\alpha}} (1-x)^{\beta}(1-xt)^{\rho}
\int^{1}_{0}dy\,\,y^{\tilde{\alpha}}(1-y)^{\beta}
(1-yt)^{\rho}
\eea
$\tilde{\alpha}=-2-\alpha-\beta-\rho.$ Next we expand the factors 
$(1-xt)^{\rho},\,(1-yt)^{\rho}$. This leads to the following form of $u^{\ast}_{1}$:
\bea
u^{\ast}_{1}&=&\gamma^{\ast}_{1}S^{\ast}_{1}\\
\gamma^{\ast}_{1}&=&\frac{\Gamma(1+\alpha')\Gamma(1+\beta')}{\Gamma(2+\alpha'+\beta')}
(\frac{\Gamma(1+\tilde{\alpha})\Gamma(1+\beta)}{\Gamma(2+\tilde{\alpha}
+\beta)})^{2}\nn\\
&=&\frac{\Gamma(\frac{7}{3}+\epsilon)\Gamma(\frac{3}{2}\epsilon)}{\Gamma(\frac{7}{3}
+\frac{5}{2}\epsilon)}(\frac{\Gamma(\frac{2}{3}+\frac{\epsilon}{2})\Gamma(2
+\frac{3}{2}\epsilon)}{\Gamma
(\frac{8}{3}+2\epsilon)})^{2}\\
S^{\ast}_{1}&=&\sum_{k}\sum_{l}\frac{(-\rho)_{k}}{k!}\frac{(-\rho)_{l}}{l!}\frac{(1+\tilde
{\alpha})_{k}}{(2+\tilde{\alpha}+\beta)_{k}}\frac{(1+\tilde{\alpha})_{l}}{(2+\tilde{\alpha}
+\beta)_{l}}\frac{(1+\alpha')_{k+l}}{(2+\alpha'+\beta')_{k+l}}\nn\\
&=&\sum_{k}\sum_{l}\frac{(\frac{4}{3}+\epsilon)_{k}}{k!}\frac{(\frac{4}{3}
+\epsilon)_{l}}{l!}\frac{(\frac{2}{3}+\frac{\epsilon}{2})_{k}}{(\frac{8}{3}
+2\epsilon)_{k}}\frac{(\frac{2}{3}+
\frac{\epsilon}{2})_{l}}{(\frac{8}{3}+2\epsilon)_{l}}\frac{(\frac{7}{3}+\epsilon)_{k+l}}
{(\frac{7}{3}+\frac{5}{2}\epsilon)_{k+l}}
\eea
One easily checks convergence of this sum.

It turns out that the contour integral $j^{(+)}_{3}$, like also
$j^{(-)}_{1}$, enters into the decomposition of the integral $I$,
eq.(B.2), with a coefficients $\sim\epsilon^{2}$.  This immediately
follows from eq.(B.2) after the substitution of the values of the
exponents
$\beta'=-1+\frac{3}{2}\epsilon,\,\,\beta=1+\frac{3}{2}\epsilon\,\,
\rho=-\frac{4}{3}-\epsilon $ eqs.(B.25)-(B.27). As a result, it is
sufficient to define the leading $\sim 1/\epsilon $ parts of the
integrals $j^{(+)}_{3},\,\,j^{(-)}_{1}$.

One finds from (B.104):
\beq
\gamma^{\ast}_{1}=\frac{2}{3\epsilon}(\frac{9}{10})^{2}+O(1)
\eeq
and from (B.105):
\beq
S^{\ast}_{1}\simeq(\sum_{k}\frac{(\frac{4}{3})_{k}(\frac{2}{3})_{k}}{k!(\frac{8}{3})_{k}})^{2}
=(\frac{\Gamma(\frac{8}{3})\Gamma(\frac{2}{3})}{\Gamma(\frac{4}{3})\Gamma(2)})^{2}=(\frac{5}{9}
)^{2}\gamma^{2}+O(\epsilon)
\eeq
This gives:
\beq
u^{\ast}_{1}=\gamma^{\ast}_{1}S^{\ast}_{1}=\frac{\gamma^{2}}{6\epsilon}+O(1)
\eeq
Finally from (B.99), one gets:
\beq
u_{3}\equiv  j^{(+)}_{3}=\frac{\gamma^{2}}{6\epsilon}-2\frac{\gamma^{2}}{6\epsilon}+
\frac{\gamma^{2}}{6\epsilon}+O(1)
\eeq
or
\beq
j^{(+)}_{3}=O(1)
\eeq
This implies that $j^{(+)}_{3}$ will not contribute to $I$ (to the finite part of $I$,
to be more precise).

Similar calculation for the integral $j^{(-)}_{1}$ gives:
\beq
j^{(-)}_{1}=-\frac{\gamma^{2}}{6\epsilon}+O(1)
\eeq

We come back now to the eq.(B.2). Substituting the values of the exponents in the
coefficients, developing the coefficients in $\epsilon $ and keeping the leading terms
only, one obtains:
\bea
I&\simeq& -\{j^{(+)}_{1}[-\pi^{3}\varepsilon^{3}j^{(-)}_{1}-\pi^{2}\varepsilon^{2}
\frac{\sqrt{3}}{2}j^{(-)}_{2}-\pi\varepsilon\frac{3}{4}j^{(-)}_{3}]\nn\\
&&\quad + j^{(+)}_{2}[-\pi^{2}\varepsilon^{2}\frac{\sqrt{3}}{2}j^{(-)}_{1}-\frac{1}{2}\pi
\varepsilon\frac{3}{4}j^{(-)}_{2}-\pi^{2}\varepsilon^{2}\frac{\sqrt{3}}{2}j^{(-)}_{3}]
\nn\\
&&\quad +j^{(+)}_{3}[\pi^{2}\varepsilon^{2}\frac{\sqrt{3}}{2}j^{(-)}_{1}-\pi^{2}
\varepsilon^{2}\frac{\sqrt{3}}{2}j^{(-)}_{2}-\pi^{3}\varepsilon^{3}j^{(-)}_{3}]\}
\eea
Here we have noted $\frac{3}{2}\epsilon=\varepsilon.$ The leading terms of the contour 
integrals are the following:
\beq
j^{(+)}_{1}\simeq\frac{\gamma^{2}}{6\epsilon},\,\,\,j^{(+)}_{2}\simeq\frac{\gamma^{2}}
{6}(\frac{1}{2}-3\kappa),\,\,\,j^{(+)}_{3}\sim 1
\eeq
\beq
j^{(-)}_{1}\simeq-\frac{\gamma^{2}}{6\epsilon},\,\,\,j^{(-)}_{2}\simeq-\frac{\gamma^{2}}
{3\epsilon},\,\,\,j^{(-)}_{3}\simeq \gamma^{2}(-\frac{7}{4}+\frac{\kappa}{2})
\eeq
Taking them into account, the expression (B.112) for $I$ can further be reduced:
\beq
I\simeq j^{(+)}_{1}(\pi^{2}\varepsilon^{2}\frac{\sqrt{3}}{2}j^{(-)}_{2}+\pi\varepsilon
\frac{3}{4}j^{(-)}_{3})+j^{(+)}_{2}\frac{1}{2}\pi\varepsilon\frac{3}{4}j^{(-)}_{2}
\eeq
It turns out finally that we need to know only the leading terms of
the integrals.  Still, the leading terms for $j^{(+)}_{2}$ and
$j^{(-)}_{3}$ are finite terms. So we had to calculate them, the
singular $\sim\frac{1}{\epsilon}$ terms for all the contour integrals
would not be sufficient.

Substituting the values (B.113), (B.114) of the integrals, replacing
back $\varepsilon$ by $\frac{3}{2}\epsilon $ and using 
$\kappa=\frac{\pi}{\sqrt{3}}+\frac{3}{2}$, one finds after
some simple algebra:
\beq
I=-\frac{\pi}{16}\gamma^{4}+O(\epsilon)
\eeq


\newpage
\small

\begin{thebibliography}{10}

\bibitem{ludwig}
A.~W.~W.~Ludwig,
\newblock {\it Nucl. Phys.}~{\bf B285}, 97 (1987);
\newblock {\it Nucl. Phys.}~{\bf B330}, 639 (1990).

\bibitem{dpp}
Vl.~S.~Dotsenko, M.~Picco and P.~Pujol,
\newblock {\it Nucl. Phys.}~{\bf B455}, 701 (1995), hep-th/9501017.

\bibitem{ddpp}
Vik.~S.~Dotsenko, Vl.~S.~Dotsenko, M.~Picco and P.~Pujol,
\newblock {\it Europhys. Lett.}~{\bf 32}, 425 (1995), hep-th/9502134.

\bibitem{ledoussal}
P.~Le Doussal and T.~Giamarchi, 
\newblock {\it Phys. Rev. Lett.}~{\bf 74}, 606 (1995), cond-mat/9409103.

\bibitem{dhss}
Vik.~S.~Dotsenko, A.~B.~Harris, D.~Sherrington and R.~B.~Stinchcombe,
{\it J. Phys.} {\bf A28}, 3093 (1995), cond-mat/9412106.

\bibitem{mpvd}
M.~M\'ezard, G.~Parisi and M.~Virasoro,
``Spin-Glass Theory and Beyond'',  World Scientific, Singapore (1987);\\
Vik.~S.~Dotsenko, ``Introduction to the Theory of Spin Glasses and
Neural Networks'', World Scientific, Singapore (1994).

\bibitem{df1}
Vl.~S.~Dotsenko and V.~A.~Fateev,
\newblock {\it Nucl. Phys.}~{\bf B240}, 312 (1984); {\bf B251}, 691 (1985).

\bibitem{df2}
Vl.~S.~Dotsenko and V.~A.~Fateev,
\newblock {\it Phys. Lett.}~{\bf 154B}, 291 (1985).

\bibitem{iw}
Y.~Imry and M.~Wortis,
\newblock {\it Phys. Rev.}~{\bf B 19}, 3581 (1979).

\bibitem{aw}
M.~Aizenman and J.~Wehr,
\newblock {\it Phys. Rev. Lett.}~{\bf 62}, 2503 (1989).

\bibitem{hb}
K.~Hui and N.~Berker,
\newblock {\it Phys. Rev. Lett.}~{\bf 62}, 2507 (1989).

\bibitem{d}
Vl.~S.~Dotsenko, {\it Advanced Studies in Pure Mathematics}~{\bf 16},
123 (1988).

\bibitem{kd} W.~Kinzel and E.~Domany,
\newblock {\it Phys. Rev.}~{\bf B 23}, 3421 (1981).

\bibitem{mp} M.~Picco, 
\newblock {\it Phys. Rev.}~{\bf B 54}, 14930 (1996), cond-mat/9507025.

\bibitem{wolf}
U.~Wolff, {\it Phys. Rev. Lett.}~{\bf 60}, 1461 (1988).

\bibitem{bd} S.~Bravy and Vik.~S.~Dotsenko (to be published).

\end{thebibliography}
\end{document}